\documentclass[aps,pre,onecolumn,superscriptaddress,amsmath,amssymb]{revtex4-2}

\usepackage[utf8]{inputenc}
\usepackage{amsmath,amsthm}
\usepackage{graphicx}
\usepackage{easyReview}
\usepackage[colorlinks = true,
linkcolor = blue,
urlcolor = blue,
citecolor = blue,
anchorcolor = blue]{hyperref}
\usepackage{braket}
\usepackage{xcolor}
\global\long\def\al{\alpha}

\global\long\def\De{\Delta}

\global\long\def\l{\lambda}

\global\long\def\d{{\mathit d}}

\global\long\def\ell#1{\theta_{#1}}
\global\long\def\bell#1{\tilde\theta_{#1}}
\global\long\def\oell#1{\bar\theta_{#1}}
\def\ir{{\mathrm i}}
\def\eE{{\mathrm e}}
\def\no{\nonumber}

\theoremstyle{remark}

\begin{document}
	
\title{A Pedestrian's Way to Baxter's Bethe Ansatz for the
    Periodic XYZ Chain}

\author{Xin Zhang}
\affiliation{Beijing National Laboratory for Condensed Matter Physics, Institute of Physics, Chinese Academy of Sciences, Beijing 100190, China}
\author{Andreas Kl\"umper}
\affiliation{Department of Physics, University of Wuppertal, Gaussstra\ss e 20, 42119 Wuppertal, Germany}
\author{Vladislav Popkov}
\affiliation{Department of Physics, University of Wuppertal, Gaussstra\ss e 20, 42119 Wuppertal, Germany}
\affiliation{Faculty of Mathematics and Physics, University of Ljubljana, Jadranska 19, SI-1000 Ljubljana, Slovenia}

\begin{abstract}
A chiral coordinate Bethe ansatz method is developed to study the periodic XYZ
chain. We construct a set of chiral vectors with fixed number of kinks.  All
vectors are factorized and have simple structures. Under roots of unity
conditions, the Hilbert space has an invariant subspace and our vectors form a
basis of this subspace.  We propose a Bethe ansatz solely based on the action of the Hamiltonian on the chiral vectors, avoiding the use of
transfer matrix techniques. This allows to parameterize the expansion coefficients and derive the homogeneous Bethe ansatz equations whose
solutions give the exact energies and eigenstates. Our analytic results agree with earlier approaches, notably by Baxter, and are supported by numerical calculations.
\end{abstract}

\maketitle

\section{Introduction}
The XYZ model, an integrable system in quantum statistical
  mechanics, has been a constant source of fascination for researchers since
  Baxter's discovery of its integrability and his groundbreaking work on the
  solution \cite{Baxter1, Baxter2, Baxter3a, Baxter3, Baxter5a, Baxter5,
    Baxter6, Baxter-book}. The exact solution for the eigenvalues of the eight
  vertex model's transfer matrix and of the Hamiltonian of the related XYZ
  quantum chain with even number of sites and periodic boundary conditions was first obtained in
  \cite{Baxter3a,Baxter-book}, then a coordinate Bethe ansatz for the
  eigenstates was presented in \cite{Baxter5a, Baxter5, Baxter6}, which was
  argued in \cite{Baxter4} to be complete.

Takhtadzhan and Faddeev successfully tackled the model using the algebraic
Bethe ansatz method \cite{Takhtadzhan1979}.  Related equations for the
eigenvectors of the eight-vertex model have been studied by \cite{FelVar96}
and by \cite{Deguchi02}.  The role of additional algebraic structures
appearing at special anisotropy parameters (roots of unity) especially in view
of the completeness of the spectrum of the transfer matrix were studied in
\cite{FabMcCoy03,FabMcCoy05,Fabricius07,FabMcCoy06,FabMcCoy09}.

The absence of a suitable vacuum state has long hindered the
application of conventional Bethe ansatz methods to the XYZ model. Exact
solutions are only attainable in specific scenarios, such as the periodic XYZ
chain with an even number of lattice sites ($N$) \cite{Baxter-book,
  Takhtadzhan1979}, or in root of unity cases or open XYZ chains
\cite{Baxter4, Takhtadzhan1979, Fan1996, Yang2006}.

The introduction of the off-diagonal Bethe ansatz method (ODBA) led to the
derivation of Bethe ansatz equations (BAE) for the spectrum of the XYZ chain
with various integrable boundary conditions \cite{Wang-book, Cao2013off,
  Cao2014}, although this approach yielded limited information about the
eigenstates.

Over the past two years, we conducted a series of studies on open XXZ and
  XYZ chains with boundary fields \cite{PhantomLong, CCBA, MPA2021,
    OpenXYZ2022}. We demonstrated the existence of two invariant subspaces in
  anisotropic Heisenberg chains under certain criteria \cite{PhantomLong,
    OpenXYZ2022}. A set of chiral vectors with kinks was constructed to expand
  the invariant subspace. Subsequently, we proposed a Bethe ansatz method to
  derive the coefficients of Bethe vectors in the chiral basis and the
  corresponding eigenvalues \cite{PhantomLong, CCBA, MPA2021, OpenXYZ2022}. A
  recent study \cite{ChiralMoreStable} shows that quantum states with helicity
  are protected from certain types of noise over intermediate timescales even
  better than the ground state, making chiral states attractive for
  experimental applications \cite{Ketterle1, Ketterle2}. Finally, chiral
  states, both in XXZ and XYZ open chains, can be targeted by
  boundary-localized strong dissipation \cite{MPA2021}.

Following our investigation of the open XYZ chain \cite{OpenXYZ2022}, we
  realized that similar invariant subspaces and chiral bases exist in the
  periodic XYZ chain. This is the motive for the present work.

In this communication, we verify that the Hilbert space of the periodic
  XYZ chain has invariant subspace(s) at roots of unity, expandable by a
  generating set of chiral vectors. Unlike the open chain, the vectors in this
  case include a free parameter, and the closure of our basis is ensured by
  the periodicity of elliptic functions at roots of unity. We employ a chiral
  coordinate Bethe ansatz to diagonalize the Hamiltonian within this
  subspace. The solutions of the resulting Bethe ansatz equations (BAE)
  determine the coefficients for the respective eigenstates in our chiral
  basis. During our calculations, we observe two distinct scenarios. When
  $M\neq\frac{N}{2}$, the subspace's dimension correlates with binomial
  coefficients, resulting in degenerate energy levels. Conversely, when
  $M=\frac{N}{2}$, we successfully construct most of the eigenstates of the
  Hamiltonian. The dimension of the invariant subspace is numerically proven
  to be equal to the number of regular solutions of the BAE. We also
  conjecture that the missing eigenstates in our chiral basis correspond to
  special ``bound pair'' BAE solutions \cite{Baxter4}. All our analytic results
  are validated through numerical checks.

Upon completing our calculations, we recognized an overlap between our
  results and the earlier work of Baxter \cite{Baxter5, Baxter6, Baxter4}. Our
  chiral basis is a subset of Baxter’s which is a truly complete basis of
  states containing two independent free parameters. All states have the
  structure of the elliptic spin helix state with insertions of kinks. The
  number of kinks is conserved under the action of the transfer matrix as
  shown in \cite{Baxter5} by the property ``pair propagation through a
  vertex'', see also \cite{Baxter-book}. Based on this property Baxter set up
  the Bethe ansatz \cite{Baxter6} for the transfer matrix. This program was
  successful in presence of two free parameters.  In contrast, our chiral
  basis and the Bethe ansatz method are based solely on the local divergence
  condition (\ref{h;psi},\ref{h;phi}) for the local Hamiltonian and the local
  vectors \cite{MPA2021, OpenXYZ2022} which are allowed to depend only on one
  free parameter instead of two. Notably, our approach is independent of the
  transfer matrix, representing a distinct methodology. While Baxter
  exclusively investigated the periodic XYZ chain, our work extends to both
  periodic and open chains. Furthermore, this paper serves as a complement to
  Baxter’s previous research. For instance, we explicitly identify the bound
  pair solutions of the Bethe ansatz equations (BAE) and correlate them with
  the missing eigenstates in our chiral subspace. The existence of the
  invariant subspace of lower dimension may facilitate the study of certain
  physical phenomena, such as quantum quenching.

The structure of this paper is as follows: We begin by revisiting the
parametrization of the XYZ model and postulating the conditions necessary for
the existence of a chiral invariant subspace.  In Section \ref{LocalStates},
we define the chiral basis vectors, which serve as the foundation of the
invariant subspace.  Subsequently, we introduce a Bethe ansatz method to
parameterize the eigenvalues and eigenstates of the Hamiltonian. We explore
specific cases, including $M=0$ and $M=1,2$, in Sections \ref{M0Case} and
\ref{M1Case}, respectively, and then proceed to generalize our findings to
arbitrary values of $M$ in Section \ref{MCase}.  In Section
\ref{sec:XXZ-XX-limits}, we delve into the XXZ and XX limits of the model.
Lastly, we provide useful identities and technical proofs in the Appendices.

\section{XYZ Model and Chiral Subspace Conditions}
\label{sec:Model}

The quantum spin-$\frac12$ XYZ chain with periodic boundary condition, defined
by the following Hamiltonian
\begin{align} 
&H=\sum_{n=1}^N\mathbf{h}_{n,n+1}=\sum_{n=1}^NJ_x\,\mathbf{\sigma}_n^x\sigma_{n+1}^x+J_y\,\sigma_n^y\sigma_{n+1}^y+J_z\,\sigma_n^z\sigma_{n+1}^z,\label{Hamiltonian}
\end{align}
is one of the most famous integrable models without $U(1)$ symmetry
\cite{Takhtadzhan1979,Baxter-book}.  Here $N$ is the length of the system and
$\sigma^x,\sigma^y,\sigma^z$ are the Pauli matrices and the periodic boundary
condition implies $\vec{\sigma}_{N+1}\equiv\vec\sigma_1$.  The exchange
coefficients $\{J_x,\,J_y,\,J_z\}$ are parameterized by the crossing parameter
$\eta$ as \cite{Yang2006,Wang-book,Cao2013off}
\begin{align}
J_x=\frac{\ell{4}(\eta)}{\ell{4}(0)},\quad J_y=\frac{\ell{3}(\eta)}{\ell{3}(0)},\quad J_z=\frac{\ell{2}(\eta)}{\ell{2}(0)},
\label{eq:AnisotropyParametrization}
\end{align}
where $\ell{\alpha}(u)\equiv\vartheta_{\al}(\pi
u,\eE^{\ir\pi\tau}),\,\alpha=1,2,3,4$ are elliptic theta functions
\cite{WatsonBook} defined in Appendix \ref{Theta;Function} and $\tau$ is a
quasi-period of $\ell{\alpha}(u)$ with ${\rm Im}[\tau]>0$.

In this paper, we study periodic XYZ chains with $\eta$ taking the following
discrete values
\begin{align}
&(N-2M)\eta =2L\tau+2K,\quad 0\leq M\leq N,\quad L,K\in\mathbb{Z}, \label{Constraint;Periodic}\\
&2(s+1)\eta=2L_0\tau+2K_0,\quad s\in\mathbb{N},\quad L_0,K_0\in\mathbb{Z},\label{Constraint;Periodic;2}
\end{align}
where $s$ is the \textit{smallest nonnegative integer} satisfying
Eq.~(\ref{Constraint;Periodic;2}).

Equation (\ref{Constraint;Periodic;2}) demands that $\eta$, $\tau$ and $1$ are commensurate. Furthermore
a canonical set of integers with smallest value for the factor of $\eta$ defines the 
non-negative integer $s$ as well as $L_0$ and $K_0$.

As (\ref{Constraint;Periodic;2}) is satisfied, Eq. (\ref{Constraint;Periodic}) may have more than the solution $M=N/2$ (for even $N$). For any
integer $M$ that satisfies (\ref{Constraint;Periodic}), we are going to set up a set of chiral states
for which a Bethe ansatz can be derived. These states contain a fixed number
$M$ of what we call kinks.  There are $s+1$ many linearly independent kink
states with same locations of the kinks.

In the following we show that the conditions (\ref{Constraint;Periodic}),
(\ref{Constraint;Periodic;2}) guarantee the existence of an invariant subspace
of the XYZ Hamiltonian, spanned by the factorized helix states with kinks, of
type shown in Fig.~\ref{Fig1}.  The number of basis states (the number of
trajectories in Fig.~\ref{Fig1} with kinks at arbitrary positions) $(s+1)
\binom{N}{M}$ typically coincides with the dimension of the invariant subspace
$d_{M,s}$ (the exception $N=2M$ will be discussed separately).  Our aim is to
find the eigenvectors of the Hamiltonian and the corresponding spectrum within
the invariant subspace.

For odd $N$, $s+1<\frac{|N-2M|}{2}$ or $s+1=|N-2M|$.  For even $N$,
$s+1\leq\frac{|N-2M|}{2}$.  It can be verified that the dimension of the
invariant subspace is strictly smaller than the Hilbert space dimension,
$d_{M,s}=(s+1) \binom{N}{M}< 2^N$.  For the exceptional case $N=2M$, Eq.
(\ref{Constraint;Periodic}) is satisfied for any $\eta$ with the choice
$K=L=0$, then, $s$ from (\ref{Constraint;Periodic;2}) can become arbitrarily
large, leading to possible $(s+1) \binom{N}{N/2}> 2^N$, and consequently a
linear dependence of states in the generating system.

\textit{Hermiticity condition.} Only when $\tau$ is purely imaginary and
$\eta$ is real or purely imaginary, the Hamiltonian is Hermitian, specifically
as follows
\begin{itemize}
\item when $\rm{Im}[\eta]=\rm{Re}[\tau]=0$,\,\, $|J_x|\geq |J_y|\geq |J_z|$, 
\item when $\rm{Re}[\eta]=\rm{Re}[\tau]=0$,\,\, $|J_x|\leq |J_y|\leq |J_z|$.
\end{itemize}

One of particular examples of a system satisfying (\ref{Constraint;Periodic}), (\ref{Constraint;Periodic;2}) 
is an XYZ spin chain on  special manifold of couplings   $J_x J_y + J_y J_z+J_z J_x=0$,  corresponding to $\eta = 2/3$ or to 
 $\eta = 2 \tau/3$
in our parametrization (\ref{eq:AnisotropyParametrization}) and discussed in detail in \cite{Hagendorf2012}.

\section{Chiral basis vectors}\label{LocalStates}
Introduce the following local ket vector \cite{Baxter6,Baxter-book}
\begin{align} 
\psi(u)=\binom{\bell{1}(u)}{-\bell{4}(u)},\label{Psi}
\end{align}
where $u\in\mathbb C$ is a free parameter and $\bell{\alpha}(u)\equiv
\vartheta_{\al}(\pi u,\eE^{2\ir\pi\tau}),\,\alpha=1,2,3,4$ are elliptic theta
functions defined in Appendix \ref{Theta;Function}. The state $\psi(u)$
possesses the quasi periodicity property:
\begin{align}
\psi(u+2k+2l\tau)=\exp\!\left[-\ir\pi l(2u+2l\tau+1)\right]\!\psi(u),\quad k,l\in\mathbb{Z}. \label{psi;periodicity}
\end{align}
The following identities  hold \cite{MPA2021,OpenXYZ2022}
\begin{align}
\mathbf h_{n,n+1}\psi_n(u)\psi_{n+1}(u\pm\eta)&=\left[\pm f(u)\,\sigma_n^z\mp f(u\pm\eta)\,\sigma_{n+1}^z+w(\pm u)\right]\psi_n(u)\psi_{n+1}(u\pm\eta),\label{h;psi}
\end{align}
where $\mathbf h_{n,n+1}$ is the local density of the XYZ Hamiltonian
(\ref{Hamiltonian}), and the functions $f(u)$, $w(u)$ are
\begin{align}
&g(u)=\frac{\ell{1}(\eta)\ell{1}'(u)}{\ell{1}'(0)\ell{1}(u)},\quad f(u)=\frac{\ell{1}(\eta)\ell{2}(u)}{\ell{2}(0)\ell{1}(u)},\label{f}\\
&w(u)=g(\eta)+g(u)-g(u+\eta).\label{w}
\end{align} 
The state $\psi(u)$ satisfies another equation \cite{OpenXYZ2022} 
\begin{align}
f(u)\,\sigma^z\psi(u)=a_\pm(u)\psi(u)+b_\pm(u)\psi(u\pm2\eta),\label{sigma;psi}
\end{align}
with
\begin{align}
a_\pm(u)=\mp\frac{\ell{2} (\eta)\ell{2}(u) \ell{1}(u\pm\eta)}{\ell{2}(0)\ell{1}(u)\ell{2}(u\pm\eta)},\qquad b_\pm(u)=\pm\frac{\ell{2}(u)}{\ell{2}(u\pm\eta)}.\label{a;b}
\end{align}
Denote 
\begin{align}
{u}_{m}=u_0+m\eta,\,\, u_0\in\mathbb{C} \label{u0}
\end{align} 
and then define the global states \cite{Baxter6}
\begin{align}
|\d;n_1,n_2,\ldots,n_M\rangle=&\bigotimes_{k_1=1}^{n_1}\psi(u_{2\d+k_1})\bigotimes_{k_2=n_1+1}^{n_2}\psi(u_{2\d+k_2-2})\cdots\bigotimes_{k_M=n_{M-1}+1}^{n_M}\psi(u_{2\d+k_M-2M+2})\no\\
&\bigotimes_{k_{M+1}=n_{M}+1}^{N}\psi(u_{2\d+k_{M+1}-2M}),\quad 1\leq n_1<n_2<\ldots<n_M\leq N.
\label{eq:Basis}
\end{align}

Using Eqs.~(\ref{Constraint;Periodic}), (\ref{Constraint;Periodic;2}),
(\ref{h;psi}) and (\ref{sigma;psi}), one can prove that
$H|\d;n_1,\ldots,n_M\rangle$ is a linear combination of
\begin{align}
|\d;n_1,\ldots,n_M\rangle\,\,\mbox{and}\,\,|\d;\ldots,n_{k-1},n_k\pm 1,n_{k+1},\ldots\rangle,\quad k=1,\ldots,M,
\end{align}
where $|\d;\ldots,n_{j},n_{j+1}=n_j,\ldots\rangle\equiv 0$.
The periodicity of the theta functions implies 
\begin{align}
&|\d;n_1,\ldots,n_{M-1},N+1\rangle\propto|\d+1;1,n_1,\ldots,n_{M-1}\rangle,\no\\
&|\d;0,n_2,\ldots,n_M\rangle\propto|\d-1;n_2,\ldots,n_M,N\rangle,\no\\
&|\d;n_1,\ldots,n_M\rangle\propto|\d+s+1;n_1,\ldots,n_M\rangle,
\end{align} 
where the last identity follows from Eq.~(\ref{Constraint;Periodic;2}).
Therefore, the vectors
\begin{align}
\ket{\d;n_1,n_2,\ldots,n_M},\quad \d=0,1,\ldots,s,\quad 1\leq n_1< n_2<\ldots<n_M\leq N,\label{Basis;M}
\end{align} form an invariant subspace of $H$.

\begin{figure}[htbp]
\includegraphics[width=0.6\textwidth]{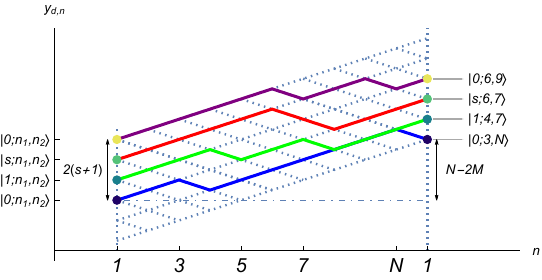}
\caption{Visualization of the vectors $\{\ket{\d;n_1,n_2}\}$ for $M=2,s=2$.
  Any state in (\ref{Basis;M}) corresponds to a directed path. Here we denote
  $\otimes_{n=1}^N\psi(\eta y_{\d,n})\equiv\ket{\d;n_1,n_2,\ldots,n_M}$.
}\label{Fig1}
\end{figure}

The qubit phase $u$ of the state (\ref{eq:Basis}) increases linearly by the
amount $\eta$ each site along the chain, except at the kink positions
$n_1,\ldots,n_M$, where it decreases by the same amount.  The chiral vectors
(\ref{Basis;M}) can be represented in a form of trajectories, see
Fig.~\ref{Fig1}, each trajectory having exactly $M$ kinks.  The total number
of kinks $M$ thus serves as a conserved charge.  In the following sections, we
diagonalize the Hamiltonian within the invariant subspace spanned by the
vectors (\ref{Basis;M}).  The total number of eigenstates thus constructed is
equal to the dimension of the invariant subspace, and provided $M\neq N/2$, it
is given by
\begin{align}
d_{M,s}=(s+1)\binom{N}{M}.
\label{eq:dMs}
\end{align}

\textit{Remark A.} ~Changing the sign of the chirality $\eta\to-\eta$, we get
another set of linearly independent basis vectors, of the same dimension as in
(\ref{eq:dMs}), provided $N\neq 2M$.  The vectors with positive and negative
chirality are all linearly independent, so that we have two invariant
subspaces, with opposite chiralities, of the dimension $d_{M,s}$ each.  Any
state obtained from the set (\ref{Basis;M}) by a shift of the initial phase is
not independent and can be expanded as a linear combination of the $2d_{M,s}$
vectors.  This is our numerical observation and for the moment we do not have
an analytical proof.

\textit{Remark B. }
~Baxter proposed a similar basis in \cite{Baxter6} which
    contains two free parameters. In comparison, there is only one free
    parameter $u_0$ in our chiral basis (\ref{Basis;M}). The generating
  system of our chiral states (\ref{Basis;M}) has a simpler structure and is
  an invariant
  subset of Baxter's states in \cite{Baxter6,Baxter4}. More details are
  given in Appendix \ref{app:Baxter}.

If the Hermiticity conditions at the end of sec.~\ref{sec:Model} are not
satisfied, the Hamiltonian (\ref{Hamiltonian}) is non-Hermitian, and the
bra-vectors need to be constructed separately.  The bra-vector analog of the
basis (\ref{Basis;M}) can be built using identities for the bra vector
$\phi(u)=\left(\bell{1}(u),\, -\bell{4}(u)\right)$ analogous to (\ref{h;psi}):
\begin{align}
&\phi_n(u)\phi_{n+1}(u\pm\eta)\mathbf h_{n,n+1}=\phi_n(u)\phi_{n+1}(u\pm\eta)\left[\pm f(u)\,\sigma_n^z\mp f(u\pm\eta)\,\sigma_{n+1}^z+w(\pm u)\right],\label{h;phi}\\
&f(u)\phi(u)\,\sigma^z=a_\pm(u)\phi(u)+b_\pm(u)\phi(u\pm2\eta).\label{sigma;phi}
\end{align}
The bra-vectors of type (\ref{Basis;M}) constituting the basis of the chiral
invariant subspace for bra vectors can then be constructed by replacement of
all $\psi_n(u)\rightarrow \phi_n(u)$ in (\ref{eq:Basis}).

\section{$M=0$ case: perfect helix without kinks} \label{M0Case}
Let us introduce the following states:
\begin{align}
\ket{\Psi^\pm_0(u)}=\bigotimes_{n=1}^N\psi(u\pm n\eta).\label{M0;eigenstate}
\end{align} 
 The state $\ket{\Psi^\pm_0(u)}$ is chiral (the sign $\pm$ in
 (\ref{M0;eigenstate}) indicates the sign of chirality) and can be viewed as
 an elliptic analog of the spin-helix eigenstate of the periodic XXZ model at
 root of unity anisotropies \cite{PhantomShort}.

Both elliptic spin-helix states (\ref{M0;eigenstate}) are eigenstates of the
XYZ Hamiltonian.  Indeed Eq.~(\ref{Constraint;Periodic}) with $M=0$ gives:
\begin{align}
\psi(u\pm N\eta)\propto\psi(u),\qquad f(u\pm N\eta)=f(u),
\end{align}
for arbitrary $u$.  Acting by the Hamiltonian on $\ket{\Psi^\pm_0(u)}$ and
using Eq.~(\ref{h;psi}), we obtain
\begin{align}
H\ket{\Psi^\pm_0(u)}=E_0\ket{\Psi^\pm_0(u)},\quad E_0
=Ng(\eta)+4\ir\pi L\,\frac{\ell{1}(\eta)}{\ell{1}'(0)}.\label{E0}
\end{align}

Due to $\bell{\al}(1-u)=\bell{\al}(u),\,\alpha=1,4$, it is easy to prove
$\ket{\Psi_0^-(u)}=\ket{\Psi^+_0(1-u)}.$ The eigenvalue $E_0$ is consistent
with the ones given by other approaches \cite{Takhtadzhan1979,Wang-book}.
Since $E_0$ is $u$-independent, the eigenvalue is degenerate.  Linearly
independent eigenstates are obtained by choosing different $u$ in
(\ref{M0;eigenstate}).  One can prove that the energy level $E_0$ is at least
$2N$-fold degenerate, see Appendix \ref{AppendixE}.  Rather remarkably, the
elliptic spin helix state (\ref{M0;eigenstate}) can be targeted in an open XYZ
chain with strong local dissipation applied to edge spins in the quantum Zeno
regime \cite{OpenXYZ2022}, the inverse strength of dissipation playing the
role of an effective temperature, and the state (\ref{M0;eigenstate}) playing
the role of an effective ground state.

\section{One-kink and two-kinks cases $M=1,2$} \label{M1Case}

\textbf{One-kink case $M=1$.}

~Using Eqs.~(\ref{Constraint;Periodic}), (\ref{Constraint;Periodic;2}) with $M=1$ we get 
\begin{align*}
f(u+N\eta-2\eta)=f(u+2(s+1)\eta)=f(u),\quad \ket{\d+s+1;n}\propto\ket{\d;n},
\end{align*}
for arbitrary $u,\d,n$.  Acting Hamiltonian $H$ on $\{|\d;n\rangle \}$ and
using Eqs.~(\ref{h;psi}) and (\ref{sigma;psi}) repeatedly, we arrive at
\begin{align}
&H|\d;n\rangle=A(u_{2\d+n})|\d;n\rangle+2A_-(u_{2\d+n})|\d;n-1\rangle+2A_+(u_{2\d+n})|\d;n+1\rangle,\quad n=1,\ldots,N-1,\no \\
&H|\d;N\rangle
={A}(u_{2\d+N})|\d;N\rangle+2A_-(u_{2\d+N})|\d;N-1\rangle+2A_+(u_{2\d+2})|\d+1;1\rangle,\no\\
&H|\d;1\rangle
=A(u_{2\d+1})|\d;1\rangle+2A_+(u_{2\d+1})|\d;2\rangle+2A_-(u_{2\d+N-1})\eE^{4\ir\pi L\eta}|\d-1;N\rangle, \label{eqM1}\\
&H\ket{s;N}
=A(u_{2s+N})|s;N\rangle+2A_-(u_{2s+N})| s;N-1\rangle+2A_+(u_{0})\eE^{4\ir\pi L_0\eta}W|0;1\rangle,\no\\
&H|0;1\rangle=A(u_{1})|0;1\rangle+2A_+(u_1)|0;2\rangle+2A_-(u_{2s+N+1})W^{-1}\eE^{4\ir\pi L\eta}|s;N\rangle,\no
\end{align}
where  $A_\pm(u),A(u)$ and  $W$ are given by
\begin{align*}	
&A_-(u)=\frac{\ell{2}(u)}{\ell{2}(u-\eta)},\quad 
A_+(u)=\frac{\ell{2}(u-\eta)}{\ell{2}(u)},\quad A_{\pm}(u+2l\tau+2k)=\eE^{\pm 4\ir\pi l\eta}A_\pm(u),\\
&A(u)=E_0+2[g(u+\tfrac12)-g(u-\eta+\tfrac12)-2g(\eta)]\\
&W=(-1)^{NL_0}\exp\left[-2\ir\pi L_0\left(2\eta+\sum_{n=1}^Nu_{n-2}+NL_0\tau\right)\right]\no\\
&=\exp\left[-2\ir\pi (sL+L+L_0)\left(2u_{s+1}+L\tau -\eta\right)-4\ir\pi L_0\eta\right].
\end{align*}
Eqs.~(\ref{eqM1}) explicitly show that the states $\ket{\d;n}$ with
$n=1,\ldots,N$ and $\d=0,1,\ldots,s$ form an invariant subspace of $H$.  The
respective chiral eigenstates are constructed as a linear combination of $\{\ket{\d;n}\}$
\begin{align}
|\Psi(\lambda)\rangle=
\sum_{\d=0}^{s}\sum_{n=1}^NF_{\d,n}(\lambda)\ket{\d;n},\quad \mbox{with}\quad H|\Psi(\l)\rangle=E(\l)|\Psi(\l)\rangle.
\label{eq:PsiM1}
\end{align}
Here we assume both the eigenstate and the energy to depend on the Bethe root
$\l$.  $E(\l)$ is parametrized as
\begin{align}
&E(\l)=E_0+E_b(\l),\label{M1;Energy}\\
&E_b(\l)=2[g(\l-\tfrac{\eta}{2})-g(\l+\tfrac{\eta}{2})].\label{Eb}
\end{align}
where $E_0$ is a ``vacuum" energy (\ref{E0}).  Substituting (\ref{eq:PsiM1})
into the eigen equation of $H$ and using (\ref{eqM1}) we obtain functional
recursive relations:
\begin{align}
&A_-(u_{2\d+n+1})F_{\d,n+1}(\l)+A_+(u_{2\d+n-1})F_{\d,n-1}(\l)=B(\l,u_{2\d+n})F_{\d,n}(\l),\no\\
&\eE^{4\ir\pi L\eta}A_-(u_{2\d+N+1})F_{\d+1,1}(\l)+A_+(u_{2\d+N-1})F_{\d,N-1}(\l)=B(\l,u_{2\d+N})F_{\d,N}(\l),\no\\
&A_+(u_{2\d})F_{\d-1,N}(\l)+A_-(u_{2\d+2})F_{\d,2}(\l)=B(\l,u_{2\d+1})F_{\d,1}(\l),\label{recursive;3}\\
&\eE^{4\ir\pi L_0\eta}WA_+(u_{0})F_{s,N}(\l)+F_{0,2}(\l)A_-(u_{2})=B(\l,u_{1})F_{0,1}(\l),\no\\
&W^{-1}\eE^{4\ir\pi L\eta}A_-(u_{2s+N+1})F_{0,1}+A_+(u_{2 s+N-1})F_{s,N-1}(\l)=B(\l,u_{2s+N})F_{s,N}(\l),\no
\end{align}
where 
\begin{align*}
B(\l,u)&=\frac{\ell{1}(\l+\frac{\eta}{2})}{\ell{1}(\l-\frac{\eta}{2})}\frac{\ell{2}(u-\eta)\ell{2}(\lambda-u-\tfrac{\eta}{2})}{\ell{2}(u)\ell{2}(\lambda-u+\tfrac{\eta}{2})}
+\frac{\ell{1}(\l-\frac{\eta}{2})}{\ell{1}(\l+\frac{\eta}{2})}\frac{\ell{2}(u)\ell{2}(\lambda-u+\tfrac{3\eta}{2})}{\ell{2}(u-\eta)\ell{2}(\lambda-u+\tfrac{\eta}{2})}.
\end{align*}

We propose the following ansatz for the coefficients $F_{\d,n}(\l)$:
\begin{align}
&F_{\d,n}(\l)=\alpha_\d\,U_{2\d+n}^{(n)}(\l),\label{M1;Ansatz}\\[2pt]
&\alpha_\d=\exp\left[2\ir\pi L \d\left(2u_{\d}+2L\tau+\eta\right)- \d\xi\right],\label{alpha}\\[2pt]
&U_{m}^{(n)}(\l)=\left[\frac{\ell{1}(\l+\frac{\eta}{2})}{\ell{1}(\l-\frac{\eta}{2})}\right]^{n}\frac{\ell{2}(\lambda-u_{m}+\frac{\eta}{2})}{\ell{2}(u_{m-1})\ell{2}(u_{m})},\label{def;U}
\end{align}
where $\xi$ is an unknown parameter. The function $U_{m}^{(n)}(\l)$ satisfies
the following identity:
\begin{align}
A_-(u_{m+1})U_{m+1}^{(n+1)}(\l)+A_+(u_{m-1})U_{m-1}^{(n-1)}(\l)=B(\l,u_{m})U_{m}^{(n)}(\l),\label{U;property}
\end{align} 
for arbitrary $\l, m, n$.  Due to (\ref{U;property}), the first of
Eqs.~(\ref{recursive;3}) is satisfied automatically.  Inserting our ansatz
(\ref{M1;Ansatz}) into the remaining Eqs.~(\ref{recursive;3}), we get the
following BAE for $\l$ and $\xi$
\begin{align}
&\left[\frac{\ell{1}(\l+\frac{\eta}{2})}{\ell{1}(\l-\frac{\eta}{2})}\right]^{N}\exp(4\ir\pi L\lambda+\xi)=1,\label{BAE;M1;1}\\
&
\exp(4\ir \pi L_0\lambda-(s+1)\xi)=1. \label{BAE;M1;2}
\end{align}

When $N$ is even, $L_0\leq |L|$. For odd $N$, $L_0<|L|$ or $L_0=|2L|$.  Our
BAE (\ref{BAE;M1;1}), (\ref{BAE;M1;2}) are consistent with the ones given by
the off-diagonal Bethe ansatz method \cite{Wang-book}.

For the Hermitian case when $\tau$ is purely imaginary and $\eta$ is real,
entailing $L=L_0=0$, our BAE (\ref{BAE;M1;1}), (\ref{BAE;M1;2}) simplify as
\begin{align}
&\eE^{\xi}\left[\frac{\ell{1}(\l+\frac{\eta}{2})}{\ell{1}(\l-\frac{\eta}{2})}\right]^{N}=1,\label{BAE;M1;3}\\
&\eE^{(s+1)\xi}=1, \label{BAE;M1;4}
\end{align}
while the coefficient $F_{\d,n}(\l)$ becomes
\begin{align*}
F_{\d,n}(\l)&=\eE^{-\d\xi}\left[\frac{\ell{1}(\l+\frac{\eta}{2})}{\ell{1}(\l-\frac{\eta}{2})}\right]^n\frac{\ell{2}(\lambda-u_{2\d+n}+\frac{\eta}{2})}{\ell{2}(u_{2\d+n-1})\ell{2}(u_{2\d+n})}.
\end{align*}

An example of BAE solutions is given in Tab.~\ref{Tab_1}.  It is
  noteworthy to point out that some of the Bethe roots in Tab.~\ref{Tab_1}
  have an obvious tight connection with $\tau$, e.g.,
  $\mbox{Im}[\l]=\frac{\tau}{2\ir}$. In the limit $\tau \rightarrow
  +\ir\infty$, the XYZ model with parameters listed in the left panel of
  Tab.~\ref{Tab_1} degenerates into the XXZ model with $\De = \cos (\pi \eta)= -
  \frac12$, (see (\ref{XXZlimit})). Some Bethe roots will tend to
  $\pm\ir\infty$, becoming so-called phantom Bethe roots, i.e.~special roots
  contributing zero energy \cite{PhantomShort} and thus leading to extra
  degeneracies in the spectrum of the XXZ Hamiltonian. For more examples of
  the sort, see Tabs.~\ref{Tab_6} and \ref{Tab_7}.

\textit{Remark A1.}~~If $\{\l,\xi\}$ is a solution of BAE (\ref{BAE;M1;1}),
(\ref{BAE;M1;2}), then $\{\l+1,\xi\}$ and $\{\l+\tau,\xi'\}$ are also BAE
solutions.  Since
\begin{align}
F_{\d,n}(\l+1)=-F_{\d,n}(\l),\quad \eE^{\xi}=\eE^{\xi'-2\ir\pi N\eta+4\ir\pi L\tau}=\eE^{\xi'-4\ir\pi \eta},\quad 
F_{\d,n}(\l+\tau)|_{\xi=\xi'}=\eE^{-2\ir\pi(\l-u_{0}+\frac{\eta}{2}+\frac{\tau}{2})}F_{\d,n}(\l),
\end{align}
these solutions are equivalent.  So we restrict the Bethe root $\l$ to the
rectangle $0\leq {\rm Re}[\l]<1,\,0\leq {\rm Im}[\l]<{\tau}/{\ir}$.

\textit{Remark B1.}~~Due to the identity $E_b(m+\tau-u)=E_b(u),\, m=0,1$, the
solutions $\{\l,\,\xi\}$ and $\{m+m'\tau-\l,\,\xi''\},\,m.m'=0,1$ correspond
to the same energy, see Tab.~\ref{Tab_1} for an example.

Next we consider the case of two-kink states.

\begin{table}[htbp]
	\centering
\begin{minipage}[c]{0.35\textwidth}
\begin{tabular}{|c|c|c|}
\hline
$\l$ & $\xi$ & $E$ \\
\hline
1.4114$\ir$ & 0 & \boxed{-4.4167} \\ 
0.0586$\ir$ & $4\ir\pi\eta$ & \boxed{-4.4167} \\
0.2025$\ir$ & 0 & $-$3.7422 \\
1.2675$\ir$ & $4\ir\pi\eta$ & $-$3.7422 \\
$\frac{\tau}{2}$ & $2\ir\pi\eta$ & $-$2.6195 \\
$\frac12$+$\frac{\tau}{2}$ & $2\ir\pi\eta$ & $-$2.3825 \\
$\frac12$+1.2283$\ir$ & 0 & $-$0.9141\\
$\frac12$+0.2417$\ir$ & $4\ir\pi\eta$ & $-$0.9141 \\
$\frac12$+0.1422$\ir$ & 0 & 0.7378\\
$\frac12$+1.3278$\ir$ & $4\ir\pi\eta$ & 0.7378 \\
$\frac12$+0.0837$\ir$ & $2\ir\pi\eta$ & 2.1776 \\
$\frac12$+1.3863$\ir$ & $2\ir\pi\eta$ & 2.1776 \\
$\frac12$+1.4307$\ir$ & 0 & 3.1550\\
$\frac12$+0.0393$\ir$ & $4\ir\pi\eta$ & 3.1550 \\
$\frac12$ & 0 & 3.5007\\
\hline
\end{tabular}
\end{minipage}
\begin{minipage}[c]{0.6\textwidth}
\begin{tabular}{|c|c|c|}
\hline
$\l$ & $\xi$ & $E$ \\
\hline
0.9209 & 1.4315$\ir$ &  \boxed{-39.0981} \\
0.0791 & $-$1.4315$\ir$ &  \boxed{-39.0981} \\
0.7417 &  $-$2.1637$\ir$ & $-$35.4475\\
0.2583 &  2.1637$\ir$ & $-$35.4475 \\
$\frac12$ &  0 &  $-$32.3420\\
$\frac12$+$\frac{\tau}{2}$ & $2\ir \pi \eta$ &  $-$9.0082 \\
0.6857+$\frac{\tau}{2}$ & $2\ir \pi \eta$$-$0.5384$\ir$ & $-$3.2905 \\
0.3143+$\frac{\tau}{2}$ & $2\ir \pi \eta$+0.5384$\ir$ & $-$3.2905 \\
0.1977+$\frac{\tau}{2}$ & $2\ir \pi \eta$+1.6559$\ir$ & 7.6856 \\
0.8023+$\frac{\tau}{2}$ & $2\ir \pi \eta$$-$1.6559$\ir$ & 7.6856 \\
0.8815+$\frac{\tau}{2}$ & $2\ir \pi \eta$$-$3.0874$\ir$ & 18.2039 \\
0.1185+$\frac{\tau}{2}$ & $2\ir \pi \eta$+3.0874$\ir$ & 18.2039 \\
0.9440+$\frac{\tau}{2}$ & $2\ir \pi \eta$+1.6253$\ir$ & 25.5295 \\ 
0.0560+$\frac{\tau}{2}$ & $2\ir \pi \eta$$-$1.6253$\ir$ & 25.5295 \\
$\frac{\tau}{2}$ & $2\ir \pi \eta$&  28.1417\\
\hline
\end{tabular}
\end{minipage}
\caption{Left: Numerical solutions of BAE (\ref{BAE;M1;3}) - (\ref{BAE;M1;4})
  with $N=5,\,M=1,\,\tau=1.47\ir$, $\eta=\frac{2}{3}$ and $s=2$. The
  exchange coefficients are $\{J_x,J_y,J_z\}=\{1.0302, 0.9710, -0.4999\}$.
  Right: Numerical solutions of BAE (\ref{BAE;M1;1}) - (\ref{BAE;M1;2}) with
  $N=5,\,M=1,\,\tau=1.47\ir$, $\eta=\frac{2\tau}{3}$ and $s=2$. The
  exchange coefficients are $\{J_x,J_y,J_z\}=\{-3.7333, 5.5542,
  11.3876\}$. Here $\boxed{\cdots}$ represents the ground state energy which
  is degenerate and belongs to the chiral invariant manifold.  We see that the
  Bethe roots $\l$ are distributed on the segment ${\rm Re}[\l]=0,\frac12$ or
  ${\rm Im}[\l]=0,\frac{\tau}{2\ir}$. Since the Hamiltonian is an even
  function of $\eta$, there exists another set of Bethe ansatz solutions with
  $\eta\to-\eta,\l\to\l,\xi\to-\xi$, which corresponds to the same set of
  energies but different eigenvectors.  Thus all the energies in any of the
  tables are degenerate. The total number of solutions is $2(s+1)N=30$.}
\label{Tab_1}
\end{table}

\bigskip

\textbf{Two-kinks case $M=2$.}

~We proceed analogously in the case $M=2$. The vectors
$\{\ket{\d;n_1,n_2}\}$ with $1\leq n_1 <n_2 \leq N$, $\d=0,\ldots, s$ form an
invariant subspace (see Appendix \ref{app:M2Case} for details).  The
corresponding eigenstates can be expanded as
\begin{align}
&|\Psi(\l_1,\l_2)\rangle=
\sum_{\d=0}^{s}\,\sum_{1\leq n_1<n_2\leq N}F_{\d,n_1,n_2}(\l_1,\l_2)\ket{\d;n_1,n_2},\label{M2;Eigenstate}\\
& H|\Psi(\l_1,\l_2)\rangle=E(\l_1,\l_2)|\Psi(\l_1,\l_2)\rangle,\label{M2;Eigenequation}\\
&E(\l_1,\l_2)=E_0+E_b(\l_1)+E_b(\l_2),\no
\end{align}
where $E_0$ and $E_b(\l)$ are defined in (\ref{E0}) and (\ref{Eb})
respectively.  Substituting Eq.~(\ref{M2;Eigenstate}) in the eigenvalue
equation (\ref{M2;Eigenequation}), we obtain functional recursive equations
for $F_{\d,n_1,n_2}$, see Appendix \ref{app:M2Case}.  The recursive equations
are solved by the following ansatz
\begin{align}
F_{\d,n_1,n_2}(\l_1,\l_2)=\alpha_\d\left[C_{1,2}\,U_{2\d+n_1}^{(n_1)}(\l_1)\,U_{2\d+n_2-2}^{(n_2)}(\l_2)+C_{2,1}\,U_{2\d+n_1}^{(n_1)}(\l_2)\,U_{2\d+n_2-2}^{(n_2)}(\l_1)\right],\label{M2;Ansatz;Main}
\end{align}
where $U_{m}^{(n)}(\l)$ is defined in (\ref{def;U}).  After some tedious
calculations, we obtain the two-body scattering matrix, see Appendix
\ref{app:Smatrix}:
\begin{align}
S_{1,2}=\frac{C_{2,1}}{C_{1,2}}=\frac{\ell{1}(\l_1-\l_2-\eta)}{\ell{1}(\l_1-\l_2+\eta)},\label{def;S}
\end{align}
and the chiral BAE for the Bethe roots $\lambda_1,\lambda_2$:
\begin{align}
&\left[\frac{\ell{1}(\l_j+\frac{\eta}{2})}{\ell{1}(\l_j-\frac{\eta}{2})}\right]^N\prod_{k\neq j}\frac{\ell{1}(\l_j-\l_k-\eta)}{\ell{1}(\l_j-\l_k+\eta)}\exp\left(4\ir\pi L\l_j+\xi\right)=1,\quad j=1,2,\label{BAE;M2;Main1}\\
&\exp[4\ir\pi L_0(\l_1+\l_2)-(s+1)\xi]=1.\label{BAE;M2;Main2}
\end{align} 
The coefficient $\alpha_\d$ in (\ref{M2;Ansatz;Main}) is parameterized in
terms of $\xi$ and $\d$ as
\begin{align}
\alpha_\d=\exp\left[2\ir\pi L \d\left(2u_{\d}+2L\tau+\eta\right)- \d\xi\right].
\end{align} 
The complete set of solutions of BAE (\ref{BAE;M2;Main1}), (\ref{BAE;M2;Main2})
for a special case is given in Tab.~\ref{N7M2;data}.

\begin{table}[htbp]
\begin{minipage}{0.5\textwidth}
\begin{tabular}{|c|c|c|c|}
\hline
$\l_1$ & $\l_2$ & $\xi$ & $E$ \\
\hline
0.9417 & 0.0220 & 1.7902i & \boxed{-11.7526} \\
0.0583 & 0.9780 & $-$1.7902i & \boxed{-11.7526} \\
0.1106 & 0.0112 & 1.0206i & $-$10.6912 \\
0.8894 & 0.9888 & $-$1.0206i & $-$10.6912 \\
0.1021 & 0.8979 & 0 & $-$9.0419 \\
0.0506 & 0.5068 & 0.4805i & $-$8.6817 \\
0.9494 & 0.4932 & $-$0.4805i & $-$8.6817 \\
0.5068+$\frac{\tau}{2}$ & 0.0506 & $2\ir \pi \eta$+0.4808i & $-$8.6706 \\
0.4932+$\frac{\tau}{2}$ & 0.9494 & $2\ir \pi \eta$$-$0.4808i & $-$8.6706 \\
0.0125 & 0.8831+$\frac{\tau}{2}$ & $2\ir \pi \eta$+1.2194i & $-$7.6183 \\
0.9875 & 0.1169+$\frac{\tau}{2}$ & $2\ir \pi \eta$$-$1.2194i & $-$7.6183 \\
0.9158+$\frac{\tau}{2}$ & 0.9715 & $2\ir \pi \eta$$-$0.9443i & $-$5.9926 \\
0.0842+$\frac{\tau}{2}$ & 0.0285 & $2\ir \pi \eta$+0.9443i & $-$5.9926 \\
$\frac12$+$\frac{\tau}{2}$ & $\frac12$ & $2\ir \pi \eta$ & $-$5.9497 \\
$\frac12$+0.0951i & $\frac12$+0.2849i & $4\ir \pi \eta$ & $-$5.9497 \\
0.1106+$\frac{\tau}{2}$ & 0.4336 & $2\ir \pi \eta$+0.3696i & $-$4.0638 \\
0.5664 & 0.8894+$\frac{\tau}{2}$ & $2\ir \pi \eta$$-$0.3696i & $-$4.0638 \\
0.5669+$\frac{\tau}{2}$ & 0.8890+$\frac{\tau}{2}$ & $4\ir \pi \eta$$-$0.3696i & $-$4.0621 \\
0.4331+$\frac{\tau}{2}$ & 0.1110+$\frac{\tau}{2}$ & $4\ir \pi \eta$+0.3696i & $-$4.0621 \\
0.9263+$\frac{\tau}{2}$ & 0.0969 & $2\ir \pi \eta$+0.1940i & $-$3.7988 \\ 0.0737+$\frac{\tau}{2}$ & 0.9031 & $2\ir \pi \eta$$-$0.1940i & $-$3.7988 \\
0.0437 & 0.9463+$\frac{\tau}{2}$ & $2\ir \pi \eta$$-$2.1785i & $-$3.2901 \\
0.9563 & 0.0537+$\frac{\tau}{2}$ & $2\ir \pi \eta$+2.1785i & $-$3.2901 \\
0.9502+0.1265i & 0.9502+0.2535i & $4\ir \pi \eta$$-$0.8350i & $-$3.2592 \\
0.0498+0.1265i & 0.0498+0.2535i & $4\ir \pi \eta$+0.8350i & $-$3.2592 \\
0.0158+0.1267i & 0.0158+0.2533i & $4\ir \pi \eta$$-$1.8298i & $-$2.6256 \\
0.9842+0.1267i & 0.9842+0.2533i & $4\ir \pi \eta$+1.8298i & $-$2.6256 \\
0.9581+$\frac{\tau}{2}$ & 0 & $2\ir \pi \eta$+1.7432i & $-$2.5499 \\
0.0419+$\frac{\tau}{2}$ & 0 & $2\ir \pi \eta$$-$1.7432i & $-$2.5499 \\
0.0392+$\frac{\tau}{2}$ & 0.0638 & $2\ir \pi \eta$+0.8629i & $-$1.2544 \\
0.9608+$\frac{\tau}{2}$ & 0.9362 & $2\ir \pi \eta$$-$0.8629i & $-$1.2544 \\
\hline 
\end{tabular}
\end{minipage}
\begin{minipage}{0.48\textwidth}
\begin{tabular}{|c|c|c|c|}
\hline 
0.9653 & 0.0153+$\frac{\tau}{2}$ & $2\ir \pi \eta$+1.9316i & 0.5804\\
0.0347 & 0.9847+$\frac{\tau}{2}$ & $2\ir \pi \eta$$-$1.9316i & 0.5804 \\
0.0190 & 0.0116+$\frac{\tau}{2}$ & $2\ir \pi \eta$$-$1.8375i & 0.5941 \\
0.9810 & 0.9884+$\frac{\tau}{2}$ & $2\ir \pi \eta$+1.8375i & 0.5941 \\
0.9767+$\frac{\tau}{2}$ & 0.0918 & $2\ir \pi \eta$+0.5736i & 1.1487 \\
0.0233+$\frac{\tau}{2}$ & 0.9082 & $2\ir \pi \eta$$-$0.5736i & 1.1487 \\
0.5063 & 0.9655+$\frac{\tau}{2}$ & $2\ir \pi \eta$$-$0.2368i & 1.6849 \\
0.4937 & 0.0345+$\frac{\tau}{2}$ & $2\ir \pi \eta$+0.2368i & 1.6849 \\
0.5063+$\frac{\tau}{2}$ & 0.9655+$\frac{\tau}{2}$ & $4\ir \pi \eta$$-$0.2369i & 1.6944 \\
0.4937+$\frac{\tau}{2}$ & 0.0345+$\frac{\tau}{2}$ & $4\ir \pi \eta$+0.2369i & 1.6944 \\
0.0493+$\frac{\tau}{2}$ & 0.8925+$\frac{\tau}{2}$ & $4\ir \pi \eta$+1.6063i & 2.0519 \\
0.9507+$\frac{\tau}{2}$ & 0.1075+$\frac{\tau}{2}$ & $4\ir \pi \eta$$-$1.6063i & 2.0519 \\
0.9158 & 0.9923+$\frac{\tau}{2}$ & $2\ir \pi \eta$$-$0.7697i & 2.1506 \\
0.0842 & 0.0077+$\frac{\tau}{2}$ & $2\ir \pi \eta$+0.7697i & 2.1506 \\
0.0703+$\frac{\tau}{2}$ & 0.9297+$\frac{\tau}{2}$ & $4\ir \pi \eta$ & 2.1783\\
$\frac12$ & $\frac{\tau}{2}$ & $2\ir \pi \eta$ & 4.2446 \\
$\frac12$+$\frac{\tau}{2}$ & $\frac{\tau}{2}$ & $4\ir \pi \eta$ & 4.2550 \\
0.0242+$\frac{\tau}{2}$ & 0.1004+$\frac{\tau}{2}$ & $4\ir \pi \eta$$-$1.0503i & 5.1168 \\
0.9758+$\frac{\tau}{2}$ & 0.8996+$\frac{\tau}{2}$ & $4\ir \pi \eta$+1.0503i & 5.1168 \\
0.1046+$\frac{\tau}{2}$ & 0.9907+$\frac{\tau}{2}$ & $4\ir \pi \eta$$-$1.2961i & 6.1318 \\
0.8954+$\frac{\tau}{2}$ & 0.0093+$\frac{\tau}{2}$ & $4\ir \pi \eta$+1.2961i & 6.1318 \\
0.0198+$\frac{\tau}{2}$ & 0.9321+$\frac{\tau}{2}$ & $4\ir \pi \eta$$-$0.4029i & 7.5138\\
0.9802+$\frac{\tau}{2}$ & 0.0679+$\frac{\tau}{2}$ & $4\ir \pi \eta$+0.4029i & 7.5138 \\
0.9864+$\frac{\tau}{2}$ & 0.9348+$\frac{\tau}{2}$ & $4\ir \pi \eta$$-$0.6600i & 8.2335 \\
0.0136+$\frac{\tau}{2}$ & 0.0652+$\frac{\tau}{2}$ & $4\ir \pi \eta$+0.6600i & 8.2335 \\
0.0469+$\frac{\tau}{2}$ & 0.9681+$\frac{\tau}{2}$ & $4\ir \pi \eta$+2.2203i & 8.2419 \\
0.9531+$\frac{\tau}{2}$ & 0.0319+$\frac{\tau}{2}$ & $4\ir \pi \eta$$-$2.2203i & 8.2419 \\
0.9552+$\frac{\tau}{2}$ & 0.9967+$\frac{\tau}{2}$ & $4\ir \pi \eta$$-$2.4978i & 10.6968 \\
0.0448+$\frac{\tau}{2}$ & 0.0033+$\frac{\tau}{2}$ & $4\ir \pi \eta$+2.4978i & 10.6968 \\
0.0069+$\frac{\tau}{2}$ & 0.9700+$\frac{\tau}{2}$ & $4\ir \pi \eta$+1.9014i & 12.2719 \\
0.9931+$\frac{\tau}{2}$ & 0.0300+$\frac{\tau}{2}$ & $4\ir \pi \eta$$-$1.9014i & 12.2719 \\
0.0178+$\frac{\tau}{2}$ & 0.9822+$\frac{\tau}{2}$ & $4\ir \pi \eta$ & 12.8134 \\
\hline 
\end{tabular}
\end{minipage}
\caption{Numerical solutions of BAE (\ref{BAE;M2;Main1}), (\ref{BAE;M2;Main2})
  with $N=7, M=2, \tau=0.38\ir$, $\eta=\frac{2\tau}{3}$ and $s=2$. The
  exchange coefficients are $\{J_x,J_y,J_z\}=\{-0.8500, 1.6986,
  1.7012\}$. The total number of solutions in the tables is
  $(s+1)\binom{N}{2}=63$.  Another $63$ solutions/eigenvectors corresponding
  to the same set of energies are obtained by a symmetry operation, i.e.
  replacement $\xi\to-\xi$ in the tables, and the replacement $\eta\to-\eta$
  in the eigenvectors' coefficients (\ref{M;Ansatz}). The total number of
  chiral solutions is $2(s+1)\binom{N}{2}=126$.  The entries of the Table
  inside boxes $\boxed{\cdots}$ represent the ground state energy.  For all
  entries, the real part of $\xi$ is given by $2 \ir \pi \eta m $, where
  $m=0,1,2$.  }
\label{N7M2;data}
\end{table}

In the next section \ref{MCase} we show, that a multi-body scattering process
between multiple kinks can be factorized as a product of two-body scatterings
governed by the chiral $S$-matrix (\ref{def;S}), which is typical for
integrable systems.  Formally, a generalization to the arbitrary $M$ case can
be done analogously to the conventional coordinate Bethe ansatz method.

\section{Generalization of the chiral Bethe Ansatz to arbitrary $M$ case}\label{MCase}
For arbitrary $M$, the eigenstates are expanded as 
\begin{align*}
&|\Psi(\l_1,\ldots,\l_M)\rangle=
\sum_{\d=0}^{s}\,\sum_{\substack{1\leq n_1<n_2<\cdots\\\cdots<n_M\leq N}}F_{\d,n_1,n_2,\ldots,n_M}(\l_1,\ldots,\l_M)\ket{\d;n_1,\ldots,n_M},
\end{align*}
where the corresponding energy is
\begin{align*}
E(\l_1,\ldots,\l_M)=E_0+\sum_{k=1}^ME_b(\l_k).
\end{align*}

We propose the following ansatz 
\begin{align}
F_{\d,n_1,n_2,\ldots,n_M}(\l_1,\ldots,\l_M)=\alpha_\d\sum_{x_1,\ldots,x_M}C_{x_1,\ldots,x_M}\prod_{k=1}^{M}U_{2\d-2k+n_k+2}^{(n_k)}(\l_{x_k}),\label{M;Ansatz}
\end{align}
where $U_{m}^{(n)}(\l)$ is defined in (\ref{def;U}) and $\{x_1,\ldots,x_M\}$
is a permutation of $\{1,\ldots,M\}$. The coefficients
$\{C_{x_1,\ldots,x_M}\}$ in terms of Bethe roots $\{\l_1,\ldots,\l_M\}$
satisfy
\begin{align}
\frac{C_{\ldots,x_{n+1},x_n,\ldots}}{C_{\ldots,x_{n},x_{n+1},\ldots}}&=\frac{\ell{1}(\l_{x_n}-\l_{x_{n+1}}-\eta)}{\ell{1}(\l_{x_n}-\l_{x_{n+1}}+\eta)}.
\end{align}
The Bethe roots and the parameter $\xi$ satisfy the following BAE
\begin{align}	&\left[\frac{\ell{1}(\l_j+\frac{\eta}{2})}{\ell{1}(\l_j-\frac{\eta}{2})}\right]^N\prod_{k\neq j}^M\frac{\ell{1}(\l_j-\l_k-\eta)}{\ell{1}(\l_j-\l_k+\eta)}\exp\left(4\ir\pi L\l_j+\xi\right)=1,\quad j=1,2,\ldots,M,\label{BAE;M;1}\\
&\exp\left(4\ir\pi L_0\sum_{k=1}^M\l_k-(s+1)\xi\right)=1.\label{BAE;M;2}
\end{align}
The coefficient $\alpha_\d$ in (\ref{M;Ansatz}) can be parameterized in terms
of $\xi$ as
\begin{align*}
\alpha_\d=\exp\left[2\ir\pi L \d\left(2u_{\d}+2L\tau+\eta\right)- \d\xi\right].
\end{align*} 

~~~
 
\textit{Remark A2.}~~One can verify that $\{\l_1,\l_2,\ldots,\l_M,\xi\}$,
$\{\l_1+1,\l_2,\ldots,\l_M,\xi\}$ and
$\{\l_1+\tau,\l_2,\ldots,\l_M,\xi+4\ir\pi\eta\}$ are equivalent solutions. If
not stated otherwise, all the Bethe roots are distinct and lie within the
rectangle $0\leq {\rm Re}[\l_j]<1$, $0\leq {\rm Im}[\l_j]<{\tau}/{\ir}$.

\textit{Remark B2.}~~Our results are independent of whether the XYZ
Hamiltonian is Hermitian or not. The only constraints are
Eqs.~(\ref{Constraint;Periodic}) and (\ref{Constraint;Periodic;2}).  We have
explicitly constructed the ket eigenstates.  With the help of
Eqs.~(\ref{h;phi}) and (\ref{sigma;phi}), we can similarly construct the
corresponding bra eigenstates.

\textit{Remark C2.}~~The Hamiltonian is invariant under the substitution
$\eta\to-\eta$.  Once $M$, $\eta$ satisfy Eq.~(\ref{Constraint;Periodic}),
another set of solutions exists with $M\to M,\,\eta\to -\eta$ (or equivalently
$M\to N-M,\,\eta\to-\eta$).  More generally, there may exist a set of integers
$\{M_1,\ldots,M_r\}$ satisfying Eq.~(\ref{Constraint;Periodic}).  In this
case, $r$ chiral invariant subspaces exist, for each of them our procedure to
construct eigenstates remains valid.

\textit{Remark D2.}~~Only part of the eigenstates (equal to the dimension of
the chiral invariant subspace) can be constructed with our method.  Note
however that the energy does not depend on the free parameter $u_0$ which
parametrizes the chiral eigenstates via (\ref{u0}), rendering some energy
levels degenerate.  The degeneracies may be related to the $\mathfrak{sl}_2$
loop algebra \cite{Deguchi02,Deguchi2001,Fabricius2001}.

\textit{Remark E2.}~~When $N$ is even and $M=\frac{N}{2}$,
Eq.~(\ref{Constraint;Periodic}) always holds and $s$ in
Eq.~(\ref{Constraint;Periodic;2}) can be arbitrary large.
For the Bethe ansatz with $M=\frac{N}{2}$, an additional selection rule for
$\{\l_1,\ldots,\l_M\}$ is necessary to ensure the eigenstate to be
non-trivial.  Based on numerics, see e.g.~Tabs. \ref{N2;data} and
\ref{N4;data}, we conjecture the following selection rule for the valid Bethe
roots
\begin{align}
2\sum_{j=1}^{M}\l_j=k+l\tau,\qquad k,l\in\mathbb{Z},\label{SelectionRule}
\end{align} 
while the solutions violating the selection rule correspond to invalid
eigenstates. The sum rule (\ref{SelectionRule}) does not hold for
$M\neq\frac{N}{2}$, which can be seen from Tabs.~\ref{Tab_1} and
\ref{N7M2;data}. The sum rule (\ref{SelectionRule}) is consistent with Baxter's observation in
\cite{Baxter4}.
\begin{table}[htbp]
\begin{minipage}{0.45\textwidth}
\begin{tabular}{|c|c|c|c|}
\hline
$\l_1$  &  $2\l_1$ & $\xi$ & $E$ \\
\hline 
0 & 0 & 0 & $-$8.3486 \\
$\frac{\tau}{2}$ & $\tau$ & $2\ir\pi\eta$ & $-$6.2176 \\
$\frac{1}{2}+\frac{\tau}{2}$ & $1+\tau$ & $2\ir\pi\eta$ & 7.0797 \\
$\frac12$ & 1 &  0 & 7.4865 \\
\hline 
\end{tabular}
\end{minipage}
\begin{minipage}{0.45\textwidth}
\begin{tabular}{|c|c|c|c|}
\hline
$\l_1$  &  $2\l_1$ & $\xi$ & $E$ \\
\hline 
0 & 0 & 0 & $-$5.5890 \\
$\frac12$ & 1 & 0 & 0.7436 \\
$\frac{1}{2}+\frac{\tau}{2}$ & $1+\tau$ & $2\ir\pi\eta$ & 0.7526 \\
$\frac{\tau}{2}$ & $\tau$ &  $2\ir\pi\eta$ & 4.0938 \\
\hline 
\end{tabular}
\end{minipage}
\caption{Left: Numerical solutions of BAE (\ref{BAE;M;1}) - (\ref{BAE;M;2})
  with $N=2, M=1, \tau=0.38\ir$, $\eta=\frac25$ and $s=4$. The exchange
  coefficients are $\{J_x,J_y,J_z\}=\{3.6415, 0.3172, 0.2155\}$.  Right:
  Numerical solutions of BAE (\ref{BAE;M;1}) - (\ref{BAE;M;2}) with
  $N=2,M=1,\tau=0.38\ir$, $\eta=\frac{2\tau}{5}$ and $s=4$. The exchange
  coefficients are $\{J_x,J_y,J_z\}=\{0.3741, 1.2093, 1.2116\}$.}
\label{N2;data}
\end{table}
\begin{table}[htbp]
\begin{minipage}{0.48\textwidth}
\begin{tabular}{|c|c|c|c|c|}
\hline
$\l_1$  & $\l_2$ & $2(\l_1\!+\!\l_2)$ & $\xi$ & $E$ \\
\hline 
0.0862$\ir$ &0.2938$\ir$& $2\tau$ &$4\ir\pi\eta$ &  $-$14.7204 \\
0 & $\frac{\tau}{2}$ & $\tau$ & $2\ir\pi\eta$ & $-$14.5662\\
0 & $\frac12$+$\frac{\tau}{2}$  & $1+\tau$ & $2\ir\pi\eta$ &  $-$1.2689\\
0 & $\frac12$ & 1& 0& $-$0.8621\\
0.8002+$\frac{3\tau}{4}$ & 0.1998+$\frac{3\tau}{4}$ & $2+3\tau$ & $6\ir\pi\eta$ & 0.0000\\
0.8002+$\frac{\tau}{4}$ & 0.1998+$\frac{\tau}{4}$ & $2+\tau$ & $2\ir\pi\eta$ & 0.0000\\
0.2903$\ir$ & $\frac12$+0.0897$\ir$ & $1+2\tau$ & $4\ir\pi\eta$ & 0.0000\\
0.0897$\ir$ &$\frac12$+0.2903$\ir$ & $1+2\tau$ & $4\ir\pi\eta$ & 0.0000\\
0.0914$\ir$ & $\frac12$+0.0986$\ir$ & $1+\tau$& $2\ir\pi\eta$ & 0.0000 \\
0.2886$\ir$ & $\frac12$+0.2814$\ir$ & $1+3\tau$& $6\ir\pi\eta$ & 0.0000 \\
--- & --- &  ---& --- & 0.0000\\
0.7994+$\frac{\tau}{2}$ & 0.2006+$\frac{\tau}{2}$& $2+2\tau$& $4\ir\pi\eta$ & 0.1486\\
$\frac{\tau}{2}$& $\frac12$+$\frac{\tau}{2}$&   $1+2\tau$ & $4\ir\pi\eta$ & 0.8621\\
$\frac12$ & $\frac{\tau}{2}$& $1+\tau$ & $2\ir\pi\eta$ & 1.2689 \\
$\frac12$ & $\frac12$+$\frac{\tau}{2}$ & $2+\tau$ &  $2\ir\pi\eta$ &  14.5662\\
$\frac12$+0.0933$\ir$ & $\frac12$+0.2867$\ir$ & $2+2\tau$ & $4\ir\pi\eta$& 14.5719\\
\hline
\end{tabular}
\end{minipage}
\begin{minipage}{0.48\textwidth}
\begin{tabular}{|c|c|c|c|c|}
\hline
$\l_1$  & $\l_2$ & $2(\l_1\!+\!\l_2)$ & $\xi$ & $E$ \\
\hline 
0.0434 & 0.9566 & 2 & 0 & $-$7.6363\\
0& $\frac12$ & 1 & 0 & $-$4.8464 \\
0 & $\frac12$+$\frac{\tau}{2}$ & $1+\tau$ & $2\ir\pi\eta$& $-$4.8374 \\
0 & $\frac{\tau}{2}$ & $\tau$ & $2\ir\pi\eta$ & $-$1.4962\\
0.8891 & 0.6109+$\frac{\tau}{2}$ & $3+\tau$& $2\ir\pi\eta$ & 0.0000\\
0.1109 & 0.3891+$\frac{\tau}{2}$ & $1+\tau$ &  $2\ir\pi\eta$ & 0.0000\\
0.8880 & 0.6120 & 3 & 0 & 0.0000 \\
0.1120 & 0.3880 & 1 & 0 & 0.0000 \\
0.9488 & 0.0512+$\frac{\tau}{2}$ &  $2+\tau$ & $2\ir\pi\eta$ & 0.0000\\
0.0512 & 0.9488+$\frac{\tau}{2}$ &  $2+\tau$ & $2\ir\pi\eta$ & 0.0000\\
---& --- & --- & --- & 0.0000 \\
$\frac12$ &$\frac12$+$\frac{\tau}{2}$ & $2+\tau$ & $2\ir\pi\eta$ & 1.4962\\
$\frac12$+0.2850$\ir$ & $\frac12$+0.0950$\ir$ & $2+2\tau$& $4\ir\pi\eta$ & 1.4962\\
$\frac12$& $\frac{\tau}{2}$ & $1+\tau$ &$2\ir\pi\eta$ & 4.8374 \\
$\frac{\tau}{2}$ & $\frac12$+$\frac{\tau}{2}$ & $1+2\tau$ &$4\ir\pi\eta$ & 
4.8464 \\
0.0622+$\frac{\tau}{2}$ & 0.9378+$\frac{\tau}{2}$ & $2+2\tau$& $4\ir\pi\eta$ & 6.1401 \\ 
\hline
\end{tabular}
\end{minipage}
\caption{Left: Numerical solutions of BAE (\ref{BAE;M;1}) - (\ref{BAE;M;2})
  with $N=4, M=2, \tau=0.38\ir$, $\eta=\frac25$ and $s=4$. The exchange
  coefficients are $\{J_x,J_y,J_z\}=\{3.6415, 0.3172, 0.2155\}$. Right:
  Numerical solutions of BAE (\ref{BAE;M;1}) - (\ref{BAE;M;2}) with
  $N=4,M=2,\tau=0.38\ir$, $\eta=\frac{2\tau}{5}$ and $s=4$. The exchange
  coefficients are $\{J_x,J_y,J_z\}=\{0.3741, 1.2093, 1.2116\}$. The missing
  eigenstate``---'' is given by $(\sigma_{1}^{-}\sigma_{2}^{-}-
  \sigma_{2}^{-}\sigma_{3}^{-}) \ket{\uparrow \uparrow \uparrow \uparrow }$.
  Note that the Bethe roots in both tables satisfy (\ref{SelectionRule}). }
\label{N4;data}
\end{table}

Let us analyze the special case $N=2M$ presented in Tab.~\ref{N4;data} in some
more detail.  The number of generating states $(s+1)\binom {N}{M} = 30$ is larger
than the dimension of the Hilbert space $2^N=16$, so one may think that all
eigenstates can be obtained.  Nevertheless, we find that among $30$ generating
states $\{\ket{\d;n,m}\}$, only $15$ are linearly independent,
  regardless of the
choice of the overall phase $u_0$ or chirality.  Any of the states
$\ket{1,2},\ket{1,4},\ket{2,3},\ket{3,4}$ where $\ket{k,l}=
\sigma_{k}^{-}\sigma_{l}^{-} \ket{\uparrow \uparrow \ldots \uparrow }$, cannot
be expanded by the states $\{\ket{\d;n,m}\}$ alone.  It can be proved that
\begin{align}
&H\ket{3,4}=H\ket{2,3}=H\ket{1,4}=H\ket{1,2}\no\\
&=(J_x-J_y)(\ket{\Omega}+\ket{1,2,3,4})+(J_x+J_y)(\ket{2,4}+\ket{1,3}),\no\\
&\ket{n_1,\ldots,n_M}=\prod_{k=1}^M\sigma_{n_k}^-\ket{\Omega},\quad \ket{\Omega}=\ket{\uparrow\uparrow\dots\uparrow}.\no
\end{align}
Therefore,
$\kappa_1\ket{3,4}+\kappa_2\ket{2,3}+\kappa_3\ket{1,4}-(\kappa_1+\kappa_2+\kappa_3)\ket{1,2}$
for any $\kappa_1,\kappa_2,\kappa_3$ is an eigenstate of $H$ with eigenvalue
$0$. We verify that $\ket{1,2}-\ket{3,4}$ and $\ket{2,3}-\ket{1,4}$ are linear
combinations of $\{\ket{\d;n,m}\}$ so that the missing eigenstate can be
constructed by letting $\kappa_2\neq-\kappa_3$. With this degree of freedom in
mind, we can choose
\begin{align}
&\ket{\Omega_1}=\ket{1,2}-\ket{2,3},  \quad H\ket{\Omega_1}=0. \label{MissingN=4}
\end{align}

For larger system size $N=2M=6$, $s>3$, again $(s+1)\binom{N}{M}>2^N$, but the
set of chiral states in (\ref{Basis;M}) contains $60$ linearly independent states,
while the Hilbert space dimension is $2^6=64$.  Denoting $\ket{k,l}'=
\sigma_{k}^{+}\sigma_{l}^{+} \ket{\downarrow \downarrow \ldots \downarrow }$,
$(\sum a_{nkl} \ket{n,k,l})' = \sum a_{nkl} \ket{n,k,l}'$, we find that the
remaining $4$ eigenstates of $H$ have a remarkably simple form:

\begin{align}
\ket{\Omega_1}&=\ket{\Phi}+\ket{\Phi}', \quad  \ket{\Omega_2}=\ket{\Phi}-\ket{\Phi}',\no\\
\ket{\Omega_3}&=\ket{\Psi}+\ket{\Psi}', \quad  \ket{\Omega_4}=\ket{\Psi}-\ket{\Psi}',\label{RemainingEigenstate;2}\\
\ket{\Phi}&=\ket{1,2}-\ket{2,3}+\ket{3,4}-\ket{4,5}+\ket{5,6}-\ket{1,6}, \no\\
\ket{\Psi}&=\ket{1,2,4}-\ket{1,4,6}-\ket{2,3,6}-\ket{2,4,5}+\ket{2,5,6}+\ket{3,4,6},\no\\
H\ket{\Omega_1}&=2(J_x-J_y+J_z)\ket{\Omega_1},\quad H\ket{\Omega_2}=-2(J_x-J_y-J_z)\ket{\Omega_2},\no\\
H\ket{\Omega_3}&=2(J_x+J_y-J_z)\ket{\Omega_3},\quad H\ket{\Omega_4}=-2(J_x+J_y+J_z)\ket{\Omega_4}.\no
\end{align}

At this point it is instructive to compare our approach with Baxter's
classical work \cite{Baxter4}. When ${\rm{Im}}[\eta]=0$, our BAE coincide with the ones obtained in
\cite{Baxter4}, where evidence for the completeness of the Bethe
  ansatz for the periodic XYZ model with even $N$ was presented.  If the BAE
solutions are complete, where do the eigenstates and the
  respective eigenvalues missing in our chiral subspace come from? We propose
that in the case $N=2M\geq 4$ the missing eigenstates are generated by ``bound
pair" solutions of BAE when two Bethe roots form a pair as
\begin{align*}
\l_1=\frac{\eta}{2},\quad\l_2=-\frac{\eta}{2},\quad\frac{\ell{1}(\l_1+\frac\eta 2)}{\ell{1}(\l_1-\frac\eta 2)}\to \infty,\quad \frac{\ell{1}(\l_2+\frac\eta 2)}{\ell{1}(\l_2-\frac\eta 2)}\to 0,\quad \frac{\ell{1}(\l_1+\frac\eta 2)}{\ell{1}(\l_1-\frac\eta 2)}\frac{\ell{1}(\l_2+\frac\eta 2)}{\ell{1}(\l_2-\frac\eta 2)}=-1,
\end{align*}
and the remaining Bethe roots $\{\l_3,\ldots,\l_{N/2}\}$ satisfy
\begin{align}
&\eE^{\xi}\,\left[\frac{\ell{1}(\l_j+\frac{\eta}{2})}{\ell{1}(\l_j-\frac{\eta}{2})}\right]^{N-1}\frac{\ell{1}(\l_j-\-\frac{3\eta}{2})}{\ell{1}(\l_j+\frac{3\eta}{2})}\,\prod_{\substack{k=3\\k\neq j}}^{N/2}\frac{\ell{1}(\l_j-\l_k-\eta)}{\ell{1}(\l_j-\l_k+\eta)}=1,\quad j=3,\ldots,\tfrac{N}{2},\\
&\exp\left(4\ir\pi L_0\sum_{k=3}^{N/2}\l_k-(s+1)\xi\right)=1.
\end{align}
A ``bound pair" $\l_1=-\l_2=\tfrac{\eta}{2}$ contributes $-4g(\eta)$ to the energy and leads to 
\begin{align}
E(\tfrac{\eta}{2},-\tfrac{\eta}{2},\l_3,\ldots,\l_{N/2})=E_0-4g(\eta)+\sum_{k=3}^{N/2}E_b(\l_k)=(N-4)g(\eta)+\sum_{k=3}^{N/2}E_b(\l_k).
\end{align}

A possibility of such bound pair solutions was envisaged in \cite{Baxter4}
and there
studied explicitly for the XXZ chain.
Here we explicitly point them out:
for $N=4$, the only ``bound pair" solution (corresponding to one ``missing"
eigenstate) is
\begin{align*}
\l_1=\tfrac{\eta}{2},\quad\l_2=-\tfrac{\eta}{2},\quad\xi=0,\quad E=0.
\end{align*}
For $N=6$, there are four ``bound pair" BAE solutions (corresponding to four
``missing" eigenstates):
\begin{align}
&(1):\l_1=\tfrac{\eta}{2},\,\,\l_2=-\tfrac{\eta}{2},\,\,\l_3=0,\,\,\xi=0,\quad E=-2(J_x+J_y+J_z),\no\\
&(2):\l_1=\tfrac{\eta}{2},\,\,\l_2=-\tfrac{\eta}{2},\,\,\l_3=\tfrac{1}{2},\,\,\xi=0,\quad E=2(J_x+J_y-J_z),\no\\
&(3):\l_1=\tfrac{\eta}{2},\,\,\l_2=-\tfrac{\eta}{2},\,\,\l_3=\tfrac{\tau}{2},\,\,\xi=2\ir\pi\eta,\quad E=-2(J_x -J_y-J_z),\no\\
&(4):\l_1=\tfrac{\eta}{2},\,\,\l_2=-\tfrac{\eta}{2},\,\,\l_3=\tfrac{1+\tau}{2},\,\,\xi=2\ir\pi\eta,\quad E=2(J_x-J_y +J_z).\no
\end{align}

When $N=8$, we find $11$ BAE solutions with bound pairs corresponding to $11$
``missing" eigenvalues, listed in Tab. ~\ref{TableN=8}.  Remarkably, the
missing eigenstates always correspond to the ``bound pair" BAE solutions. We
conjecture that this correspondence holds for larger systems with $N=2M\geq
10$ as well.  For $N=2M=4,6,8,10,12$, the total number of missing eigenstates
(i.e.~not expandable by our set of chiral states) is $1,4,11,37,66$
respectively, according to a numerical analysis, provided that $d_{M,s}$
(\ref{eq:dMs}) is larger than $2^N$.  The relative fraction of missing
eigenstates $\frac{1}{16},\frac{4}{2^6}, \frac{11}{2^{8}}, \frac{37}{2^{10}},
\frac{66}{2^{12}}$ for $N=4,6,8,10,12$ is getting smaller with the system
size.

Summarising, we propose that the standard BAEs for periodic XYZ and even $N$
contain two types of solutions: the regular ones (constituting the vast
majority), and those with a bound pair $\l_1=-\l_2=\tfrac{\eta}{2}$.
Remarkably, all eigenstates corresponding to regular BAE solutions can be
understood in terms of our simple chiral basis with kinks (\ref{Basis;M}), see
also Fig.~\ref{Fig1}.  Unlike regular eigenstates, the ``bound pair"
eigenstates seem to lack such an appealing representation, although we can
readily construct them, using the generalized basis from \cite{Baxter6}, see
Appendix \ref{Basis;Baxter}.  Explicit expressions for the ``bound pair"
eigenstates (\ref{MissingN=4}), (\ref{RemainingEigenstate;2}) suggest that
actually these eigenstates might have a simpler representation in the usual
computational basis.

Further technical details concerning ``bound pair" eigenstates, are given in
Appendix \ref{Basis;Baxter}.

Our findings readily reduce to the partially anisotropic Heisenberg Hamiltonian,
see Section \ref{sec:XXZ-XX-limits}.

\begin{table}[htbp]
\begin{tabular}{|c|c|c|c|c|c|c|}
\hline
~$\l_1$~ & ~$\l_2$~  & $\l_3$ & $\l_4$ &$2\sum_{k}\l_k$ & $\xi$  & $E$ \\
\hline
$\frac{\eta}{2}$& $-\frac{\eta}{2}$& 0.0424i & 0.3376i & $2\tau$ & $4\ir\pi\eta$ & $-$16.1265 \\
$\frac{\eta}{2}$& $-\frac{\eta}{2}$ &$\frac{\tau}{2}$ & 0 & $\tau$ & $2\ir\pi\eta$ & $-$14.5662 \\
$\frac{\eta}{2}$& $-\frac{\eta}{2}$ & 0.1366i &0.2434i & $2\tau$ & $4\ir\pi\eta$ & $-$13.1249 \\
$\frac{\eta}{2}$& $-\frac{\eta}{2}$ &$\frac12$+$\frac{\tau}{2}$ & 0 & $1+\tau$ & $2\ir\pi\eta$ & $-$1.2689 \\
$\frac{\eta}{2}$& $-\frac{\eta}{2}$ & $\frac12$ & 0 & 1 & 0 & $-$0.8621 \\
$\frac{\eta}{2}$& $-\frac{\eta}{2}$& 0.8000+$\frac{\tau}{2}$ & 0.2000+$\frac{\tau}{2}$ & $2+2\tau$ &$4\ir\pi\eta$ & 0.0759 \\
$\frac{\eta}{2}$& $-\frac{\eta}{2}$& $\frac{\tau}{2}$ & $\frac12$+$\frac{\tau}{2}$ & $1+2\tau$ & $4\ir\pi\eta$ & 0.8621 \\
$\frac{\eta}{2}$& $-\frac{\eta}{2}$ & $\frac12$ & $\frac{\tau}{2}$ & $1+\tau$ & $2\ir\pi\eta$ & 1.2689 \\
$\frac{\eta}{2}$& $-\frac{\eta}{2}$ & $\frac12$+0.2425$\ir$ & $\frac12$+0.1375$\ir$ & $2+2\tau$& $4\ir\pi\eta$ & 14.2999 \\
$\frac{\eta}{2}$& $-\frac{\eta}{2}$ &$\frac12$+$\frac{\tau}{2}$ & $\frac12$ & $2+\tau$& $2\ir\pi\eta$ & 14.5662 \\
$\frac{\eta}{2}$& $-\frac{\eta}{2}$ &$\frac12$+0.0422i & $\frac12$+0.3378i & $2+2\tau$& $4\ir\pi\eta$ & 14.8756 \\
\hline
\end{tabular}
\caption{``Bound pair" solutions of BAE (\ref{BAE;M;1}) - (\ref{BAE;M;2}) with
  $N=8, M=4, \tau=0.38\ir$, $\eta=\frac25$ and $s=4$. The exchange coefficients are $\{J_x,J_y,J_z\}=\{3.6415, 0.3172, 0.2155\}$. }
\label{TableN=8}
\end{table}

\section{XXZ and XX limits of the XYZ model}
\label{sec:XXZ-XX-limits}

By letting $\tau\to+\ir\infty$ the XYZ chain degenerates into the partially
anisotropic XXZ chain as
\begin{align}
&J_x\to1,\quad J_y\to1,\quad J_z\to\cos(\pi\eta). \label{XXZlimit}
\end{align}
In the limit $u\to\tilde u+\frac{\tau}{2}$, $\tau\to+\ir\infty$ ($\tilde u$
being finite), we get
\begin{align*}
&\lim_{\substack{\tau\to+\ir\infty\\u\to\tilde u+\frac{\tau}{2}}}\frac{\ell{2}(u)}{\ell{1}(u)}=-\ir,\quad
\lim_{\substack{\tau\to+\ir\infty\\u\to\tilde u+\frac{\tau}{2}}}\frac{\bell{4}(u)}{\bell{1}(u)}=-\eE^{\ir\pi(\tilde u+\frac{1}{2})},\no\\
&\lim_{\substack{\tau\to+\ir\infty\\u\to\tilde u+\frac{\tau}{2}}}f(u)=-\ir\sin(\pi\eta),\quad\lim_{\substack{\tau\to+\ir\infty\\u\to\tilde u+\frac{\tau}{2}}}w(u)=\cos(\pi\eta),
\end{align*} 
The divergence condition (\ref{h;psi}) degenerates into the XXZ divergence
condition \cite{PhantomLong,PhantomShort,CCBA,SHS_Dynamic}
\begin{align}
\mathbf h_{n,n+1}^{\rm XXZ}\tilde\psi_n(\tilde u)\tilde\psi_{n+1}(\tilde u\pm\eta)&=\left[\mp\ir\sin(\pi\eta)\,\sigma_n^z\pm\ir\sin(\pi\eta)\,\sigma_{n+1}^z+\cos(\pi\eta)\right]\tilde\psi_n(\tilde u)\tilde\psi_{n+1}(\tilde u\pm\eta),
\end{align}
where $\mathbf h^{\rm XXZ}_{n,n+1}=\sigma_n^x\sigma_{n+1}^x+\sigma_n^y\sigma_{n+1}^y+\cos(\pi\eta)\,\sigma_n^z\sigma_{n+1}^z$ and $\tilde\psi(u)=(1,\,\eE^{\ir\pi( u+\frac{1}{2})})^T$.

The parameter $\eta$ now can only take some discrete real values according to
Eqs.~(\ref{Constraint;Periodic}), (\ref{Constraint;Periodic;2})
\begin{align}
&(N-2M)\eta =2K,\quad 0\leq M\leq N,\quad K\in\mathbb{Z}, \no\\
&2(s+1)\eta=2K_0,\quad s\in\mathbb{N},\quad K_0\in\mathbb{Z}.\label{Constraint;XXZ}
\end{align}
Eqs.~(\ref{Constraint;XXZ}) are constraints under which a chiral invariant
subspace exists for a periodic XXZ spin-$\frac12$ chain.  Thus we can follow
the same procedure to study the periodic XXZ chain at roots of unity
\cite{Braak2001,Deguchi2001,Fabricius2001}. The corresponding BAE obtained
from our chiral coordinate Bethe ansatz are deformed ones
\cite{PhantomShort,Braak2001}
\begin{align}
&\eE^{\bar\xi}\,\left[\frac{\sinh(\bar\l_j+\frac{\bar\eta}{2})}{\sinh(\bar\l_j-\frac{\bar\eta}{2})}\right]^N\prod_{k\neq j}^M\frac{\sinh(\bar\l_j-\bar\l_k-\bar\eta)}{\sinh(\bar\l_j-\bar\l_k+\bar\eta)}=1,\quad \eE^{(s+1)\bar\xi}=1,\quad\bar\eta=\ir\pi\eta,\quad j=1,\ldots,M.\label{BAE;XXZ}
\end{align}
From Eq.~(\ref{Constraint;XXZ}), we find the parameter $\bar\xi$ in
  (\ref{BAE;XXZ}) can take values $\pm2\bar m\bar\eta,\,\bar
  m\in\mathbb{Z}$. It implies that the conventional BAE for the periodic XXZ chain
  at root of unity (\ref{Constraint;XXZ}) may have solutions with $\bar m_0$
  phantom Bethe roots and $M$ regular Bethe roots \cite{PhantomShort}, and
  $\pm$ represents different chirality. In the six-vertex model limit $\tau\to
  +\ir\infty$, the Bethe roots in Eqs.~(\ref{BAE;M;1}), (\ref{BAE;M;2}) will
  degenerate into the ones for the XXZ model in (\ref{BAE;XXZ}) as
\begin{align}
\lim_{\tau\to+\ir\infty}\ir\pi\l_j=\bar\l_j,\quad  \lim_{\tau\to+\ir\infty}\xi=\bar\xi.\label{XYZ;XXZ}
\end{align} 
If $\ir\,{\mbox{Im}}[\l_j]/\tau$ tends toward a finite but non-zero number in
the limit $\tau\to+\ir\infty$, the corresponding Bethe root $\bar\l_j$ will be
a phantom one with $\mbox{Re}[\bar\l_j]\to\pm\infty$ \cite{PhantomShort}.  To
demonstrate this phenomenon, we do some simple numerical calculations for the
$M=1, 2$ cases with details shown in Tabs. \ref{Tab_6} and \ref{Tab_7}.

The eigenstates constructed by using the chiral basis with fixed number of
kinks are not eigenstates of the magnetization operator. However, in the XXZ
limit one may project such an eigenstate onto any of the subspaces with fixed
magnetization. The result is either 0 or an eigenstate of the Hamiltonian with
well defined magnetization. In this way several eigenstates with same energy,
but different magnetizations, resp. different numbers of Bethe rapidities in
the conventional Bethe ansatz are obtained. The degeneracy of the energy
eigenvalue in the chiral Bethe ansatz is seen by the different choices of the
parameter $u_0$ that enters the eigenstate, but not the eigenvalue expression.

\begin{table}[htbp]
	\centering
\begin{minipage}[c]{0.45\textwidth}
\begin{tabular}{|c|c|c|}
\hline
$\l$ & ${\xi}/{(2\ir\pi\eta)}$ & $E$ \\
\hline
$-$0.0586i & 1 & $-$4.4131 \\
0.0586i & $-1$ & $-$4.4131 \\
0.2027i & 0 & $-$3.7368 \\
$-$0.2027i & 0 & $-$3.7368 \\
$\frac{\tau}{2}$ & 1 & $-$2.5421 \\
$\frac12-\frac{\tau}{2}$ & $-1$ & $-$2.4581 \\
$\frac12-$0.2420i & 1 & $-$0.9176 \\
$\frac12$+0.2420i & $-1$ & $-$0.9176\\
$\frac12$+0.1423i & 0 & 0.7363 \\
$\frac12-$0.1423i & 0 & 0.7363 \\
$\frac12$+0.0837i & 1 & 2.1767 \\
$\frac12-$0.0837i & $-1$ & 2.1767 \\
$\frac12-$0.0393i & 1 & 3.1543 \\
$\frac12$+0.0393i & $-1$ & 3.1543 \\
$\frac12$ & 0 & 3.5001 \\
\hline
\end{tabular}
\end{minipage}
\begin{minipage}[c]{0.45\textwidth}
\begin{tabular}{|c|c|c|}
\hline
$\bar\l$ & $\bar\xi/(2\bar\eta)$ & $E$ \\
\hline
0.1841 & 1 & $-$4.4126 \\
$-$0.1841 & $-1$ & $-$4.4126 \\
0.6369 & 0 & $-$3.7361 \\
$-$0.6369 & 0 & $-$3.7361 \\
$-\infty$ & 1 & $-$2.5000 \\
$\infty+\frac{\ir\pi}{2}$ & $-1$ & $-$2.5000 \\
0.7602+$\frac{\ir\pi}{2}$ & 1 & $-$0.9181\\
$-$0.7602+$\frac{\ir\pi}{2}$ & $-1$ & $-$0.9181 \\
$-$0.4470+$\frac{\ir\pi}{2}$ & 0 & 0.7361 \\
0.4470+$\frac{\ir\pi}{2}$ & 0 & 0.7361 \\
$-$0.2630+$\frac{\ir\pi}{2}$ & 1 & 2.1765 \\
0.26295+$\frac{\ir\pi}{2}$ & $-1$ & 2.1765 \\
0.1233+$\frac{\ir\pi}{2}$ & 1 & 3.1542 \\
$-$0.1233+$\frac{\ir\pi}{2}$ & $-1$ & 3.1542 \\
$\frac{\ir\pi}{2}$ & 0 & 3.5000 \\
\hline
\end{tabular}
\end{minipage}
\caption{Left: Numerical solutions of BAE (\ref{BAE;M;1}) - (\ref{BAE;M;2})
  with $N=5,\,M=1,\,\tau=1.8\ir$, $\eta=\frac{2}{3}$ and $s=2$. The exchange
  coefficients are $\{J_x,J_y,J_z\}=\{1.0106, 0.9896, -0.5000\}$. Right:
  Numerical solutions of BAE (\ref{BAE;XXZ}) with $N=5,\,M=1$,
  $\eta=\frac{2}{3}$ and $s=2$. The exchange coefficients are
  $\{J_x,J_y,J_z\}=\{1, 1, -\frac12\}$. Here we let the Bethe root $\l$ lie
  within a proper range in order to compare the solutions of BAE
  for XYZ and XXZ model in a better way. Although the parameter $\tau$ we
  chose is not too large, one can still see the correspondence in
  (\ref{XYZ;XXZ}) approximately. We can verify that BAE in (\ref{BAE;XXZ}) has
  two phantom solutions.}
\label{Tab_6}
\end{table}

\begin{table}[htbp]
\begin{minipage}{0.45\textwidth}
\begin{tabular}{|c|c|c|c|}
\hline 
$\l_1$ & $\l_2$ & $\xi/(2\ir\pi\eta)$ & $E$\\
\hline 
0.1141i & $-$0.1141i & 0 & $-6.3093$ \\
0 & $\frac{\tau}{2}$ & 1 & $-4.0510$ \\
0 & $\frac12-\frac{\tau}{2}$ & $-1$ & $-3.9497$ \\
 0 & $\frac12$ & 0 & $-1.2360$ \\
$-$0.1347i & $\frac12$+0.1347i & 0 & 0.0000\\
0.1347i & $\frac12-$0.1347i & 0 & 0.0000 \\
$-$0.2764i & $\frac12+$  $\frac{\tau}{2}+$0.2764i & $1$ & 0.0000 \\
0.2764i & $\frac12- \frac{\tau}{2}-$0.2764i  & $-1$ & 0.0000 \\
 0.3193i & $\frac{\tau}{2}-$0.3193i  & 1 & 0.0000 \\
 $-$0.3193i & $-\frac{\tau}{2}+$0.3193i & $-1$ & 0.0000 \\
 $\frac{\eta}{2}$& $-\frac{\eta}{2}$ & 0 & \boxed{0.0000} \\
 0.2486+$\frac{\tau}{2}$ & 0.7514+$\frac{\tau}{2}$ & 2 & 1.2356 \\
 $\frac12-\frac{\tau}{2}$ & $-\frac{\tau}{2}$ & $-2$ & 1.2360 \\
 $\frac{\tau}{2}$ & $\frac12$ & 1 & 3.9497 \\
 $\frac12-\frac{\tau}{2}$ & $\frac12$ & $-1$ & 4.0510 \\
 $\frac12-$0.1637i & $\frac12$+0.1637i & 0 & 5.0737 \\
\hline
\end{tabular}
\end{minipage}
\begin{minipage}{0.45\textwidth}
\begin{tabular}{|c|c|c|c|}
\hline 
$\bar\l_1$ & $\bar\l_2$ & $\bar\xi/(2\bar\eta)$ & $E$\\
\hline
$-$0.3584 & 0.3584& 0 & $-$6.3085 \\
0 & $-\infty$ & 1 & $-$4.0000 \\
0 & $\infty$+$\frac{\ir \pi }{2}$ & $-1$ & $-$4.0000 \\
0 & $\frac{\ir \pi }{2}$ & 0 & $-$1.2361 \\
0.4232 & $-$0.4232+$\frac{\ir\pi}{2}$ & 0 & 0.0000 \\
$-$0.4232 & 0.4232+$\frac{\ir\pi}{2}$ & 0 & 0.0000 \\
0.9214 & $-\infty$+$\frac{\ir\pi}{2}$ & $1$ & 0.0000 \\
$-$0.9214 & $\infty$+$\frac{\ir \pi }{2}$ & $-1$ & 0.0000 \\
$-$0.9214 & $-\infty$ & 1 & 0.0000 \\
0.9214 & $\infty$ & $-1$ & 0.0000 \\
$\frac{\bar\eta}{2}$& $-\frac{\bar\eta}{2}$ & 0 & \boxed{0.0000}  \\
$-\infty$+$\frac{\ir \pi }{4}$ & $-\infty$+$\frac{3\ir \pi }{4}$ & 2 & 1.2361 \\
$\infty$ & $\infty$+$\frac{\ir \pi }{2}$ & $-2$ & 1.2361 \\
$-\infty$ & $\frac{\ir \pi }{2}$ & 1 & 4.0000 \\
$\infty$+$\frac{\ir \pi }{2}$ & $\frac{\ir \pi}{2}$ & $-1$ & 4.0000 \\
0.5142+$\frac{\ir\pi}{2}$ & $-$0.5142+$\frac{\ir\pi}{2}$ & 0 & 5.0725 \\
\hline
\end{tabular}
\end{minipage}
\caption{Left: Numerical solutions of BAE (\ref{BAE;M;1}) - (\ref{BAE;M;2})
  with $N=4,\,M=2,\,\tau=1.8\ir$, $\eta=\frac{2}{5}$ and $s=4$. The
  exchange coefficients are $\{J_x,J_y,J_z\}=\{1.0128, 0.9874,
  0.3090\}$. Here the Bethe roots  satisfy the sum rule in
  (\ref{SelectionRule}). Right: Numerical solutions of BAE (\ref{BAE;XXZ})
  with $N=4,\,M=2$, $\eta=\frac{2}{5}$ and $s=4$. The exchange coefficients
  are $\{J_x,J_y,J_z\}=\{1, 1, 0.3090\}$. Here ``---'' represent the bound
  pair solution. One can see a lot of phantom solutions in the XXZ case. In
  some solutions, both $\bar\l_1$ and $\bar\l_2$ are phantom and they form a
  string. Such phantom strings will appear in the $M\geq2$ cases. Here $\boxed{\cdots}$ represents the missing ``bound pair" solution.}
\label{Tab_7}
\end{table}

For $\eta=0$, the XYZ model degenerates into the isotropic XXX model.  In the
isotropic case, the spin helical structure disappears and our basis vectors
become indistinguishable. The corresponding eigenstates can now be constructed
by the conventional Bethe ansatz with the number of Bethe roots $M$ ranging from
$0$ to $N/2$ and by use of $su(2)$ operators.

Another interesting case is the XX model ($\eta=\frac12$) which can be
transformed to a free fermion model via the Jordan-Wigner transformation. From
Eqs.~(\ref{Constraint;Periodic}), (\ref{Constraint;Periodic;2}), our chiral
generating states work for the XX model with an even $N$, specifically as follows
\cite{ChiralBasisXX}
\begin{align}
&N=4m,\quad s=1, \quad M=0,2,\ldots,N,\quad m\in\mathbb{N^+},\no\\
&N=4m+2,\quad s=1, \quad M=1,3,\ldots,N-1,\quad m\in\mathbb{N}.\no
\end{align}
Under the condition $\eta=\frac{1}{2}$, the following identity holds:
$\psi(u+2\eta)=-\sigma^z\psi(u)$, rendering the basis vectors with the
argument shift $2\eta=1$ linearly independent and orthogonal to the
non-shifted ones.  Remarkably, the joined set of ``shifted" and original
chiral basis vectors for even $N$ turns out to be orthonormal and complete
\cite{ChiralBasisXX}, the latter feature resulting from
\begin{align}
&2\sum_{k=0}^{2m}\binom{N}{2k}=2^N,\quad\mbox{if}\,\, N=4m,\quad m\in\mathbb{N^+}, \no\\
&2\sum_{k=0}^{2m}\binom{N}{2k+1}=2^N,\quad\mbox{if}\,\, N=4m+2,\quad m\in\mathbb{N}.\no
\end{align}

\section*{Conclusion}
In this paper, we studied the periodic XYZ chain under the conditions
(\ref{Constraint;Periodic}) and (\ref{Constraint;Periodic;2}), which is the
elliptic analog of the root of unity conditions for anisotropy of the
partially anisotropic XXZ case.  Under these conditions, a chiral subspace
invariant under the action of the Hamiltonian exists, consisting of chiral
vectors with a fixed number of kinks.  We propose a coordinate Bethe ansatz to
find all eigenstates corresponding to the chiral invariant manifold.  The
solution of the homogeneous BAE in (\ref{BAE;M;1}), (\ref{BAE;M;2}) for the
Bethe roots gives the eigenvalues and generates the coefficients of the
respective eigenstates in the chiral basis.  This renders the chiral basis
more natural for the diagonalization problem of the XYZ chain than the
standard computational basis.

We used two parameterizations of the XYZ chain with real and also with
imaginary values of $\eta$.  Remarkably, the obtained BAE for the case of real
valued $\eta$ coincide with those obtained by earlier alternative methods
\cite{Baxter4,Takhtadzhan1979,Wang-book}.  In case of imaginary values of
$\eta$ we obtain BAE with non-root of unity twist factors. By use of the
conjugate modulus transformation this can be reconciled completely with the
results obtained in Baxter's parameterization using real values for $\eta$.

The solutions of these BAE for our invariant chiral subspace (and $M=N/2$) are
regular solutions: the states outside the invariant subspace are given by
Bethe root distributions containing a bound pair which appear in Baxter's treatment.

The integer $M$ in Eq.~(\ref{Constraint;Periodic}) denotes the number of
kinks in a chiral invariant subspace and serves as a quantum number, which should
correspond to a certain symmetry.  It is challenging to understand the origin
of this symmetry and the resulting degeneracies of the energy levels.  More
work in this directions needs to be done.

Notably,  in some of our examples (Tables 6 and 7) we treat both the XYZ chain and its XXZ counterpart (arizing as
the XYZ limit for $\tau \rightarrow \ir \infty$), demonstrating  the ``XYZ" origin of phantom Bethe roots in periodic XXZ chain  \cite{PhantomShort}. The XXZ limit is of additional interest since its open chain version with appropriately chosen boundary fields describes a
paradigmatic classical stochastic system (ASEP)
where the usage of chiral states was also found to be beneficial,  see \cite{Schuetz2023} and the references therein.

Another interesting open question is how to construct the eigenstates of the
XYZ model with generic integrable boundary conditions. Based on the known
inhomogeneous $T$-$Q$ relation, a reasonable approach would be to retrieve the
Bethe-type eigenstate via Sklyanin's separation of variables (SoV) method
\cite{BetheStateXXZ,ZhangRetrieve}.

Finally, we like to note that recent experiments in cold atoms
\cite{Ketterle1,Ketterle2} showed a feasible way to create
spin-helix states: the partially anisotropic counterparts of the states
forming our chiral invariant subspace basis.

\section*{Acknowledgments}
X.~Z.~acknowledges financial support from the National Natural Science
Foundation of China (No.~12204519). X.~Z.~thanks Y.~Wang, J.~Cao and
W.-L.~Yang for valuable discussions.
A.~K. and V.~P. acknowledge financial support from the Deutsche
Forschungsgemeinschaft through DFG project KL 645/20-2.


\begin{thebibliography}{40}%
\makeatletter
\providecommand \@ifxundefined [1]{%
 \@ifx{#1\undefined}
}%
\providecommand \@ifnum [1]{%
 \ifnum #1\expandafter \@firstoftwo
 \else \expandafter \@secondoftwo
 \fi
}%
\providecommand \@ifx [1]{%
 \ifx #1\expandafter \@firstoftwo
 \else \expandafter \@secondoftwo
 \fi
}%
\providecommand \natexlab [1]{#1}%
\providecommand \enquote  [1]{``#1''}%
\providecommand \bibnamefont  [1]{#1}%
\providecommand \bibfnamefont [1]{#1}%
\providecommand \citenamefont [1]{#1}%
\providecommand \href@noop [0]{\@secondoftwo}%
\providecommand \href [0]{\begingroup \@sanitize@url \@href}%
\providecommand \@href[1]{\@@startlink{#1}\@@href}%
\providecommand \@@href[1]{\endgroup#1\@@endlink}%
\providecommand \@sanitize@url [0]{\catcode `\\12\catcode `\$12\catcode
  `\&12\catcode `\#12\catcode `\^12\catcode `\_12\catcode `\%12\relax}%
\providecommand \@@startlink[1]{}%
\providecommand \@@endlink[0]{}%
\providecommand \url  [0]{\begingroup\@sanitize@url \@url }%
\providecommand \@url [1]{\endgroup\@href {#1}{\urlprefix }}%
\providecommand \urlprefix  [0]{URL }%
\providecommand \Eprint [0]{\href }%
\providecommand \doibase [0]{https://doi.org/}%
\providecommand \selectlanguage [0]{\@gobble}%
\providecommand \bibinfo  [0]{\@secondoftwo}%
\providecommand \bibfield  [0]{\@secondoftwo}%
\providecommand \translation [1]{[#1]}%
\providecommand \BibitemOpen [0]{}%
\providecommand \bibitemStop [0]{}%
\providecommand \bibitemNoStop [0]{.\EOS\space}%
\providecommand \EOS [0]{\spacefactor3000\relax}%
\providecommand \BibitemShut  [1]{\csname bibitem#1\endcsname}%
\let\auto@bib@innerbib\@empty
\bibitem [{\citenamefont {Baxter}(1971{\natexlab{a}})}]{Baxter1}%
  \BibitemOpen
  \bibfield  {author} {\bibinfo {author} {\bibfnamefont {R.~J.}\ \bibnamefont
  {Baxter}},\ }\bibfield  {title} {\bibinfo {title} {Eight-vertex model in
  lattice statistics},\ }\href
  {https://link.aps.org/doi/10.1103/PhysRevLett.26.832} {\bibfield  {journal}
  {\bibinfo  {journal} {Phys. Rev. Lett.}\ }\textbf {\bibinfo {volume} {26}},\
  \bibinfo {pages} {832} (\bibinfo {year} {1971}{\natexlab{a}})}\BibitemShut
  {NoStop}%
\bibitem [{\citenamefont {Baxter}(1971{\natexlab{b}})}]{Baxter2}%
  \BibitemOpen
  \bibfield  {author} {\bibinfo {author} {\bibfnamefont {R.}~\bibnamefont
  {Baxter}},\ }\bibfield  {title} {\bibinfo {title} {One-dimensional
  anisotropic heisenberg chain},\ }\href
  {https://link.aps.org/doi/10.1103/PhysRevLett.26.834} {\bibfield  {journal}
  {\bibinfo  {journal} {Phys. Rev. Lett.}\ }\textbf {\bibinfo {volume} {26}},\
  \bibinfo {pages} {834} (\bibinfo {year} {1971}{\natexlab{b}})}\BibitemShut
  {NoStop}%
\bibitem [{\citenamefont {Baxter}(1972{\natexlab{a}})}]{Baxter3a}%
  \BibitemOpen
  \bibfield  {author} {\bibinfo {author} {\bibfnamefont {R.~J.}\ \bibnamefont
  {Baxter}},\ }\bibfield  {title} {\bibinfo {title} {Partition function of the
  eight-vertex lattice model},\ }\href
  {https://doi.org/10.1016/0003-4916(72)90335-1} {\bibfield  {journal}
  {\bibinfo  {journal} {Ann. Phys.}\ }\textbf {\bibinfo {volume} {70}},\
  \bibinfo {pages} {193} (\bibinfo {year} {1972}{\natexlab{a}})}\BibitemShut
  {NoStop}%
\bibitem [{\citenamefont {Baxter}(1972{\natexlab{b}})}]{Baxter3}%
  \BibitemOpen
  \bibfield  {author} {\bibinfo {author} {\bibfnamefont {R.~J.}\ \bibnamefont
  {Baxter}},\ }\bibfield  {title} {\bibinfo {title} {One-dimensional
  anisotropic heisenberg chain},\ }\href
  {https://doi.org/10.1016/0003-4916(72)90270-9} {\bibfield  {journal}
  {\bibinfo  {journal} {Ann. Phys.}\ }\textbf {\bibinfo {volume} {70}},\
  \bibinfo {pages} {323} (\bibinfo {year} {1972}{\natexlab{b}})}\BibitemShut
  {NoStop}%
\bibitem [{\citenamefont {Baxter}(1973{\natexlab{a}})}]{Baxter5a}%
  \BibitemOpen
  \bibfield  {author} {\bibinfo {author} {\bibfnamefont {R.}~\bibnamefont
  {Baxter}},\ }\bibfield  {title} {\bibinfo {title} {Eight-vertex model in
  lattice statistics and one-dimensional anisotropic heisenberg chain. i. some
  fundamental eigenvectors},\ }\href
  {https://doi.org/10.1016/0003-4916(73)90439-9} {\bibfield  {journal}
  {\bibinfo  {journal} {Ann. Phys.}\ }\textbf {\bibinfo {volume} {76}},\
  \bibinfo {pages} {1} (\bibinfo {year} {1973}{\natexlab{a}})}\BibitemShut
  {NoStop}%
\bibitem [{\citenamefont {Baxter}(1973{\natexlab{b}})}]{Baxter5}%
  \BibitemOpen
  \bibfield  {author} {\bibinfo {author} {\bibfnamefont {R.}~\bibnamefont
  {Baxter}},\ }\bibfield  {title} {\bibinfo {title} {Eight-vertex model in
  lattice statistics and one-dimensional anisotropic heisenberg chain. ii.
  equivalence to a generalized ice-type lattice model},\ }\href
  {https://doi.org/10.1016/0003-4916(73)90440-5} {\bibfield  {journal}
  {\bibinfo  {journal} {Ann. Phys.}\ }\textbf {\bibinfo {volume} {76}},\
  \bibinfo {pages} {25} (\bibinfo {year} {1973}{\natexlab{b}})}\BibitemShut
  {NoStop}%
\bibitem [{\citenamefont {Baxter}(1973{\natexlab{c}})}]{Baxter6}%
  \BibitemOpen
  \bibfield  {author} {\bibinfo {author} {\bibfnamefont {R.}~\bibnamefont
  {Baxter}},\ }\bibfield  {title} {\bibinfo {title} {Eight-vertex model in
  lattice statistics and one-dimensional anisotropic heisenberg chain. iii.
  eigenvectors of the transfer matrix and hamiltonian},\ }\href
  {https://doi.org/10.1016/0003-4916(73)90441-7} {\bibfield  {journal}
  {\bibinfo  {journal} {Ann. Phys.}\ }\textbf {\bibinfo {volume} {76}},\
  \bibinfo {pages} {48} (\bibinfo {year} {1973}{\natexlab{c}})}\BibitemShut
  {NoStop}%
\bibitem [{\citenamefont {Baxter}(1982)}]{Baxter-book}%
  \BibitemOpen
  \bibfield  {author} {\bibinfo {author} {\bibfnamefont {R.~J.}\ \bibnamefont
  {Baxter}},\ }\href@noop {} {\emph {\bibinfo {title} {{Exactly Solved Models
  in Statistical Mechanics}}}}\ (\bibinfo  {publisher} {Academic Press},\
  \bibinfo {year} {1982})\BibitemShut {NoStop}%
\bibitem [{\citenamefont {Baxter}(2002)}]{Baxter4}%
  \BibitemOpen
  \bibfield  {author} {\bibinfo {author} {\bibfnamefont {R.~J.}\ \bibnamefont
  {Baxter}},\ }\bibfield  {title} {\bibinfo {title} {{Completeness of the Bethe
  ansatz for the six and eight-vertex models}},\ }\href
  {https://doi.org/10.1023/A:1015437118218} {\bibfield  {journal} {\bibinfo
  {journal} {J. Stat. Phys.}\ }\textbf {\bibinfo {volume} {108}},\ \bibinfo
  {pages} {1} (\bibinfo {year} {2002})}\BibitemShut {NoStop}%
\bibitem [{\citenamefont {Takhtadzhan}\ and\ \citenamefont
  {Faddeev}(1979)}]{Takhtadzhan1979}%
  \BibitemOpen
  \bibfield  {author} {\bibinfo {author} {\bibfnamefont {L.~A.}\ \bibnamefont
  {Takhtadzhan}}\ and\ \bibinfo {author} {\bibfnamefont {L.~D.}\ \bibnamefont
  {Faddeev}},\ }\bibfield  {title} {\bibinfo {title} {{The quantum method of
  the inverse problem and the Heisenberg XYZ model}},\ }\href
  {https://doi.org/10.1070/rm1979v034n05abeh003909} {\bibfield  {journal}
  {\bibinfo  {journal} {Rush. Math. Surveys}\ }\textbf {\bibinfo {volume}
  {34}},\ \bibinfo {pages} {11} (\bibinfo {year} {1979})}\BibitemShut {NoStop}%
\bibitem [{\citenamefont {Felder}\ and\ \citenamefont
  {Varchenko}(1996)}]{FelVar96}%
  \BibitemOpen
  \bibfield  {author} {\bibinfo {author} {\bibfnamefont {G.}~\bibnamefont
  {Felder}}\ and\ \bibinfo {author} {\bibfnamefont {A.}~\bibnamefont
  {Varchenko}},\ }\bibfield  {title} {\bibinfo {title} {Algebraic bethe ansatz
  for the elliptic quantum group $e_{\tau,\eta}(sl_2)$},\ }\href@noop {}
  {\bibfield  {journal} {\bibinfo  {journal} {Nucl. Phys. B}\ }\textbf
  {\bibinfo {volume} {480}},\ \bibinfo {pages} {485–503} (\bibinfo {year}
  {(1996})}\BibitemShut {NoStop}%
\bibitem [{\citenamefont {Deguchi}(2002)}]{Deguchi02}%
  \BibitemOpen
  \bibfield  {author} {\bibinfo {author} {\bibfnamefont {T.}~\bibnamefont
  {Deguchi}},\ }\bibfield  {title} {\bibinfo {title} {Construction of some
  missing eigenvectors of the xyz spin chain at the discrete coupling constants
  and the exponentially large spectral degeneracy of the transfer matrix},\
  }\href@noop {} {\bibfield  {journal} {\bibinfo  {journal} {J. Phys. A: Math.
  Gen.}\ }\textbf {\bibinfo {volume} {35}},\ \bibinfo {pages} {879} (\bibinfo
  {year} {2002})}\BibitemShut {NoStop}%
\bibitem [{\citenamefont {Fabricius}\ and\ \citenamefont
  {McCoy}(2003)}]{FabMcCoy03}%
  \BibitemOpen
  \bibfield  {author} {\bibinfo {author} {\bibfnamefont {K.}~\bibnamefont
  {Fabricius}}\ and\ \bibinfo {author} {\bibfnamefont {B.~M.}\ \bibnamefont
  {McCoy}},\ }\bibfield  {title} {\bibinfo {title} {New developments in the
  eight vertex model. journal of statistical physics},\ }\href
  {https://doi.org/10.1023/A:1022213209641} {\bibfield  {journal} {\bibinfo
  {journal} {J. Stat. Phys.}\ }\textbf {\bibinfo {volume} {111}},\ \bibinfo
  {pages} {323–337} (\bibinfo {year} {2003})}\BibitemShut {NoStop}%
\bibitem [{\citenamefont {Fabricius}\ and\ \citenamefont
  {McCoy}(2005)}]{FabMcCoy05}%
  \BibitemOpen
  \bibfield  {author} {\bibinfo {author} {\bibfnamefont {K.}~\bibnamefont
  {Fabricius}}\ and\ \bibinfo {author} {\bibfnamefont {B.~M.}\ \bibnamefont
  {McCoy}},\ }\bibfield  {title} {\bibinfo {title} {New developments in the
  eight vertex model ii. chains of odd length},\ }\href
  {https://doi.org/10.1007/s10955-005-4410-5} {\bibfield  {journal} {\bibinfo
  {journal} {J. Stat. Phys.}\ }\textbf {\bibinfo {volume} {120}},\ \bibinfo
  {pages} {37–70} (\bibinfo {year} {2005})}\BibitemShut {NoStop}%
\bibitem [{\citenamefont {Fabricius}(2007)}]{Fabricius07}%
  \BibitemOpen
  \bibfield  {author} {\bibinfo {author} {\bibfnamefont {K.}~\bibnamefont
  {Fabricius}},\ }\bibfield  {title} {\bibinfo {title} {A new q-matrix in the
  eight-vertex model},\ }\href {https://doi.org/10.1088/1751-8113/40/15/002}
  {\bibfield  {journal} {\bibinfo  {journal} {J. Phys. A: Math. Theor.}\
  }\textbf {\bibinfo {volume} {40}},\ \bibinfo {pages} {4075} (\bibinfo {year}
  {2007})}\BibitemShut {NoStop}%
\bibitem [{\citenamefont {Fabricius}\ and\ \citenamefont
  {McCoy}(2006)}]{FabMcCoy06}%
  \BibitemOpen
  \bibfield  {author} {\bibinfo {author} {\bibfnamefont {K.}~\bibnamefont
  {Fabricius}}\ and\ \bibinfo {author} {\bibfnamefont {B.~M.}\ \bibnamefont
  {McCoy}},\ }\bibfield  {title} {\bibinfo {title} {An elliptic current
  operator for the eight-vertex model},\ }\href
  {https://doi.org/10.1088/0305-4470/39/48/003} {\bibfield  {journal} {\bibinfo
   {journal} {J. Phys. A: Math. Gen.}\ }\textbf {\bibinfo {volume} {39}},\
  \bibinfo {pages} {14869} (\bibinfo {year} {2006})}\BibitemShut {NoStop}%
\bibitem [{\citenamefont {Fabricius}\ and\ \citenamefont
  {McCoy}(2009)}]{FabMcCoy09}%
  \BibitemOpen
  \bibfield  {author} {\bibinfo {author} {\bibfnamefont {K.}~\bibnamefont
  {Fabricius}}\ and\ \bibinfo {author} {\bibfnamefont {B.~M.}\ \bibnamefont
  {McCoy}},\ }\bibfield  {title} {\bibinfo {title} {New q matrices and their
  functional equations for the eight vertex model at elliptic roots of unity},\
  }\href {https://doi.org/10.1007/s10955-009-9692-6} {\bibfield  {journal}
  {\bibinfo  {journal} {J. Stat. Phys.}\ }\textbf {\bibinfo {volume} {134}},\
  \bibinfo {pages} {643–668} (\bibinfo {year} {2009})}\BibitemShut {NoStop}%
\bibitem [{\citenamefont {Fan}\ \emph {et~al.}(1996)\citenamefont {Fan},
  \citenamefont {Hou}, \citenamefont {Shi},\ and\ \citenamefont
  {Yang}}]{Fan1996}%
  \BibitemOpen
  \bibfield  {author} {\bibinfo {author} {\bibfnamefont {H.}~\bibnamefont
  {Fan}}, \bibinfo {author} {\bibfnamefont {B.-Y.}\ \bibnamefont {Hou}},
  \bibinfo {author} {\bibfnamefont {K.-J.}\ \bibnamefont {Shi}},\ and\ \bibinfo
  {author} {\bibfnamefont {Z.-X.}\ \bibnamefont {Yang}},\ }\bibfield  {title}
  {\bibinfo {title} {{Algebraic Bethe ansatz for the eight-vertex model with
  general open boundary conditions}},\ }\href
  {https://doi.org/10.1016/0550-3213(96)00398-7} {\bibfield  {journal}
  {\bibinfo  {journal} {Nucl. Phys. B}\ }\textbf {\bibinfo {volume} {478}},\
  \bibinfo {pages} {723} (\bibinfo {year} {1996})}\BibitemShut {NoStop}%
\bibitem [{\citenamefont {Yang}\ and\ \citenamefont {Zhang}(2006)}]{Yang2006}%
  \BibitemOpen
  \bibfield  {author} {\bibinfo {author} {\bibfnamefont {W.-L.}\ \bibnamefont
  {Yang}}\ and\ \bibinfo {author} {\bibfnamefont {Y.-Z.}\ \bibnamefont
  {Zhang}},\ }\bibfield  {title} {\bibinfo {title} {{T--Q relation and exact
  solution for the XYZ chain with general non-diagonal boundary terms}},\
  }\href {https://doi.org/10.1016/j.nuclphysb.2006.03.025} {\bibfield
  {journal} {\bibinfo  {journal} {Nucl. Phys. B}\ }\textbf {\bibinfo {volume}
  {744}},\ \bibinfo {pages} {312} (\bibinfo {year} {2006})}\BibitemShut
  {NoStop}%
\bibitem [{\citenamefont {Wang}\ \emph {et~al.}(2016)\citenamefont {Wang},
  \citenamefont {Yang}, \citenamefont {Cao},\ and\ \citenamefont
  {Shi}}]{Wang-book}%
  \BibitemOpen
  \bibfield  {author} {\bibinfo {author} {\bibfnamefont {Y.}~\bibnamefont
  {Wang}}, \bibinfo {author} {\bibfnamefont {W.-L.}\ \bibnamefont {Yang}},
  \bibinfo {author} {\bibfnamefont {J.}~\bibnamefont {Cao}},\ and\ \bibinfo
  {author} {\bibfnamefont {K.}~\bibnamefont {Shi}},\ }\href@noop {} {\emph
  {\bibinfo {title} {{Off-Diagonal Bethe Ansatz for Exactly Solvable
  Models}}}}\ (\bibinfo  {publisher} {Springer},\ \bibinfo {year}
  {2016})\BibitemShut {NoStop}%
\bibitem [{\citenamefont {Cao}\ \emph {et~al.}(2013)\citenamefont {Cao},
  \citenamefont {Yang}, \citenamefont {Shi},\ and\ \citenamefont
  {Wang}}]{Cao2013off}%
  \BibitemOpen
  \bibfield  {author} {\bibinfo {author} {\bibfnamefont {J.}~\bibnamefont
  {Cao}}, \bibinfo {author} {\bibfnamefont {W.-L.}\ \bibnamefont {Yang}},
  \bibinfo {author} {\bibfnamefont {K.}~\bibnamefont {Shi}},\ and\ \bibinfo
  {author} {\bibfnamefont {Y.}~\bibnamefont {Wang}},\ }\bibfield  {title}
  {\bibinfo {title} {{Off-diagonal Bethe ansatz solutions of the anisotropic
  spin-1/2 chains with arbitrary boundary fields}},\ }\href
  {https://doi.org/10.1016/j.nuclphysb.2013.10.001} {\bibfield  {journal}
  {\bibinfo  {journal} {Nucl. Phys. B}\ }\textbf {\bibinfo {volume} {877}},\
  \bibinfo {pages} {152} (\bibinfo {year} {2013})}\BibitemShut {NoStop}%
\bibitem [{\citenamefont {Cao}\ \emph {et~al.}(2014)\citenamefont {Cao},
  \citenamefont {Cui}, \citenamefont {Yang}, \citenamefont {Shi},\ and\
  \citenamefont {Wang}}]{Cao2014}%
  \BibitemOpen
  \bibfield  {author} {\bibinfo {author} {\bibfnamefont {J.}~\bibnamefont
  {Cao}}, \bibinfo {author} {\bibfnamefont {S.}~\bibnamefont {Cui}}, \bibinfo
  {author} {\bibfnamefont {W.-L.}\ \bibnamefont {Yang}}, \bibinfo {author}
  {\bibfnamefont {K.}~\bibnamefont {Shi}},\ and\ \bibinfo {author}
  {\bibfnamefont {Y.}~\bibnamefont {Wang}},\ }\bibfield  {title} {\bibinfo
  {title} {{Spin-1/2 XYZ model revisit: General solutions via off-diagonal
  Bethe ansatz}},\ }\href
  {https://doi.org/https://doi.org/10.1016/j.nuclphysb.2014.06.026} {\bibfield
  {journal} {\bibinfo  {journal} {Nucl. Phys. B}\ }\textbf {\bibinfo {volume}
  {886}},\ \bibinfo {pages} {185} (\bibinfo {year} {2014})}\BibitemShut
  {NoStop}%
\bibitem [{\citenamefont {Zhang}\ \emph
  {et~al.}(2021{\natexlab{a}})\citenamefont {Zhang}, \citenamefont
  {Kl\"umper},\ and\ \citenamefont {Popkov}}]{PhantomLong}%
  \BibitemOpen
  \bibfield  {author} {\bibinfo {author} {\bibfnamefont {X.}~\bibnamefont
  {Zhang}}, \bibinfo {author} {\bibfnamefont {A.}~\bibnamefont {Kl\"umper}},\
  and\ \bibinfo {author} {\bibfnamefont {V.}~\bibnamefont {Popkov}},\
  }\bibfield  {title} {\bibinfo {title} {{Phantom Bethe roots in the integrable
  open spin-$\frac{1}{2}$ XXZ chain}},\ }\href
  {https://doi.org/10.1103/PhysRevB.103.115435} {\bibfield  {journal} {\bibinfo
   {journal} {Phys. Rev. B}\ }\textbf {\bibinfo {volume} {103}},\ \bibinfo
  {pages} {115435} (\bibinfo {year} {2021}{\natexlab{a}})}\BibitemShut
  {NoStop}%
\bibitem [{\citenamefont {Zhang}\ \emph
  {et~al.}(2021{\natexlab{b}})\citenamefont {Zhang}, \citenamefont
  {Kl\"umper},\ and\ \citenamefont {Popkov}}]{CCBA}%
  \BibitemOpen
  \bibfield  {author} {\bibinfo {author} {\bibfnamefont {X.}~\bibnamefont
  {Zhang}}, \bibinfo {author} {\bibfnamefont {A.}~\bibnamefont {Kl\"umper}},\
  and\ \bibinfo {author} {\bibfnamefont {V.}~\bibnamefont {Popkov}},\
  }\bibfield  {title} {\bibinfo {title} {{Chiral coordinate Bethe ansatz for
  phantom eigenstates in the open XXZ spin-$\frac{1}{2}$ chain}},\ }\href
  {https://doi.org/10.1103/PhysRevB.104.195409} {\bibfield  {journal} {\bibinfo
   {journal} {Phys. Rev. B}\ }\textbf {\bibinfo {volume} {104}},\ \bibinfo
  {pages} {195409} (\bibinfo {year} {2021}{\natexlab{b}})}\BibitemShut
  {NoStop}%
\bibitem [{\citenamefont {Popkov}\ \emph {et~al.}(2022)\citenamefont {Popkov},
  \citenamefont {Zhang},\ and\ \citenamefont {Prosen}}]{MPA2021}%
  \BibitemOpen
  \bibfield  {author} {\bibinfo {author} {\bibfnamefont {V.}~\bibnamefont
  {Popkov}}, \bibinfo {author} {\bibfnamefont {X.}~\bibnamefont {Zhang}},\ and\
  \bibinfo {author} {\bibfnamefont {T.}~\bibnamefont {Prosen}},\ }\bibfield
  {title} {\bibinfo {title} {{Boundary-driven XYZ chain: Inhomogeneous
  triangular matrix product ansatz}},\ }\href
  {https://link.aps.org/doi/10.1103/PhysRevB.105.L220302} {\bibfield  {journal}
  {\bibinfo  {journal} {Phys. Rev. B}\ }\textbf {\bibinfo {volume} {105}},\
  \bibinfo {pages} {L220302} (\bibinfo {year} {2022})}\BibitemShut {NoStop}%
\bibitem [{\citenamefont {Zhang}\ \emph {et~al.}(2022)\citenamefont {Zhang},
  \citenamefont {Kl{\"u}mper},\ and\ \citenamefont {Popkov}}]{OpenXYZ2022}%
  \BibitemOpen
  \bibfield  {author} {\bibinfo {author} {\bibfnamefont {X.}~\bibnamefont
  {Zhang}}, \bibinfo {author} {\bibfnamefont {A.}~\bibnamefont {Kl{\"u}mper}},\
  and\ \bibinfo {author} {\bibfnamefont {V.}~\bibnamefont {Popkov}},\
  }\bibfield  {title} {\bibinfo {title} {Invariant subspaces and elliptic
  spin-helix states in the integrable open spin-$\frac12$ xyz chain},\ }\href
  {https://link.aps.org/doi/10.1103/PhysRevB.106.075406} {\bibfield  {journal}
  {\bibinfo  {journal} {Phys. Rev. B}\ }\textbf {\bibinfo {volume} {106}},\
  \bibinfo {pages} {075406} (\bibinfo {year} {2022})}\BibitemShut {NoStop}%
\bibitem [{\citenamefont {K\"uhn}\ \emph {et~al.}(2023)\citenamefont {K\"uhn},
  \citenamefont {Gerken}, \citenamefont {Funcke}, \citenamefont {Hartung},
  \citenamefont {Stornati}, \citenamefont {Jansen},\ and\ \citenamefont
  {Posske}}]{ChiralMoreStable}%
  \BibitemOpen
  \bibfield  {author} {\bibinfo {author} {\bibfnamefont {S.}~\bibnamefont
  {K\"uhn}}, \bibinfo {author} {\bibfnamefont {F.}~\bibnamefont {Gerken}},
  \bibinfo {author} {\bibfnamefont {L.}~\bibnamefont {Funcke}}, \bibinfo
  {author} {\bibfnamefont {T.}~\bibnamefont {Hartung}}, \bibinfo {author}
  {\bibfnamefont {P.}~\bibnamefont {Stornati}}, \bibinfo {author}
  {\bibfnamefont {K.}~\bibnamefont {Jansen}},\ and\ \bibinfo {author}
  {\bibfnamefont {T.}~\bibnamefont {Posske}},\ }\bibfield  {title} {\bibinfo
  {title} {Quantum spin helices more stable than the ground state: Onset of
  helical protection},\ }\href {https://doi.org/10.1103/PhysRevB.107.214422}
  {\bibfield  {journal} {\bibinfo  {journal} {Phys. Rev. B}\ }\textbf {\bibinfo
  {volume} {107}},\ \bibinfo {pages} {214422} (\bibinfo {year}
  {2023})}\BibitemShut {NoStop}%
\bibitem [{\citenamefont {Jepsen}\ \emph {et~al.}(2021)\citenamefont {Jepsen},
  \citenamefont {Ho}, \citenamefont {Amato-Grill}, \citenamefont {Dimitrova},
  \citenamefont {Demler},\ and\ \citenamefont {Ketterle}}]{Ketterle1}%
  \BibitemOpen
  \bibfield  {author} {\bibinfo {author} {\bibfnamefont {P.~N.}\ \bibnamefont
  {Jepsen}}, \bibinfo {author} {\bibfnamefont {W.~W.}\ \bibnamefont {Ho}},
  \bibinfo {author} {\bibfnamefont {J.}~\bibnamefont {Amato-Grill}}, \bibinfo
  {author} {\bibfnamefont {I.}~\bibnamefont {Dimitrova}}, \bibinfo {author}
  {\bibfnamefont {E.}~\bibnamefont {Demler}},\ and\ \bibinfo {author}
  {\bibfnamefont {W.}~\bibnamefont {Ketterle}},\ }\bibfield  {title} {\bibinfo
  {title} {Transverse spin dynamics in the anisotropic heisenberg model
  realized with ultracold atoms},\ }\href
  {https://link.aps.org/doi/10.1103/PhysRevX.11.041054} {\bibfield  {journal}
  {\bibinfo  {journal} {Phys. Rev. X}\ }\textbf {\bibinfo {volume} {11}},\
  \bibinfo {pages} {041054} (\bibinfo {year} {2021})}\BibitemShut {NoStop}%
\bibitem [{\citenamefont {Jepsen}\ \emph {et~al.}(2022)\citenamefont {Jepsen},
  \citenamefont {Lee}, \citenamefont {Lin}, \citenamefont {Dimitrova},
  \citenamefont {Margalit}, \citenamefont {Ho},\ and\ \citenamefont
  {Ketterle}}]{Ketterle2}%
  \BibitemOpen
  \bibfield  {author} {\bibinfo {author} {\bibfnamefont {P.~N.}\ \bibnamefont
  {Jepsen}}, \bibinfo {author} {\bibfnamefont {Y.~K.}\ \bibnamefont {Lee}},
  \bibinfo {author} {\bibfnamefont {H.}~\bibnamefont {Lin}}, \bibinfo {author}
  {\bibfnamefont {I.}~\bibnamefont {Dimitrova}}, \bibinfo {author}
  {\bibfnamefont {Y.}~\bibnamefont {Margalit}}, \bibinfo {author}
  {\bibfnamefont {W.~W.}\ \bibnamefont {Ho}},\ and\ \bibinfo {author}
  {\bibfnamefont {W.}~\bibnamefont {Ketterle}},\ }\bibfield  {title} {\bibinfo
  {title} {Long-lived phantom helix states in heisenberg quantum magnets},\
  }\href {https://doi.org/10.1038/s41567-022-01651-7} {\bibfield  {journal}
  {\bibinfo  {journal} {Nature Physics}\ }\textbf {\bibinfo {volume} {18}},\
  \bibinfo {pages} {899} (\bibinfo {year} {2022})}\BibitemShut {NoStop}%
\bibitem [{\citenamefont {Whittaker}\ and\ \citenamefont
  {Watson}(1950)}]{WatsonBook}%
  \BibitemOpen
  \bibfield  {author} {\bibinfo {author} {\bibfnamefont {E.~T.}\ \bibnamefont
  {Whittaker}}\ and\ \bibinfo {author} {\bibfnamefont {G.~N.}\ \bibnamefont
  {Watson}},\ }\href@noop {} {\emph {\bibinfo {title} {A course of modern
  analysis}}}\ (\bibinfo  {publisher} {Cambridge University Press},\ \bibinfo
  {year} {1950})\BibitemShut {NoStop}%
\bibitem [{\citenamefont {Hagendorf}\ and\ \citenamefont
  {Fendley}(2012)}]{Hagendorf2012}%
  \BibitemOpen
  \bibfield  {author} {\bibinfo {author} {\bibfnamefont {C.}~\bibnamefont
  {Hagendorf}}\ and\ \bibinfo {author} {\bibfnamefont {P.}~\bibnamefont
  {Fendley}},\ }\bibfield  {title} {\bibinfo {title} {The eight-vertex model
  and lattice supersymmetry},\ }\href
  {https://doi.org/10.1007/s10955-012-0430-0} {\bibfield  {journal} {\bibinfo
  {journal} {JOURNAL OF STATISTICAL PHYSICS}\ }\textbf {\bibinfo {volume}
  {146}},\ \bibinfo {pages} {1122} (\bibinfo {year} {2012})}\BibitemShut
  {NoStop}%
\bibitem [{\citenamefont {Popkov}\ \emph {et~al.}(2021)\citenamefont {Popkov},
  \citenamefont {Zhang},\ and\ \citenamefont {Kl\"umper}}]{PhantomShort}%
  \BibitemOpen
  \bibfield  {author} {\bibinfo {author} {\bibfnamefont {V.}~\bibnamefont
  {Popkov}}, \bibinfo {author} {\bibfnamefont {X.}~\bibnamefont {Zhang}},\ and\
  \bibinfo {author} {\bibfnamefont {A.}~\bibnamefont {Kl\"umper}},\ }\bibfield
  {title} {\bibinfo {title} {{Phantom Bethe excitations and spin helix
  eigenstates in integrable periodic and open spin chains}},\ }\href
  {https://doi.org/10.1103/PhysRevB.104.L081410} {\bibfield  {journal}
  {\bibinfo  {journal} {Phys. Rev. B}\ }\textbf {\bibinfo {volume} {104}},\
  \bibinfo {pages} {L081410} (\bibinfo {year} {2021})}\BibitemShut {NoStop}%
\bibitem [{\citenamefont {Deguchi}\ \emph {et~al.}(2001)\citenamefont
  {Deguchi}, \citenamefont {Fabricius},\ and\ \citenamefont
  {McCoy}}]{Deguchi2001}%
  \BibitemOpen
  \bibfield  {author} {\bibinfo {author} {\bibfnamefont {T.}~\bibnamefont
  {Deguchi}}, \bibinfo {author} {\bibfnamefont {K.}~\bibnamefont {Fabricius}},\
  and\ \bibinfo {author} {\bibfnamefont {B.~M.}\ \bibnamefont {McCoy}},\
  }\bibfield  {title} {\bibinfo {title} {The sl2 loop algebra symmetry of the
  six-vertex model at roots of unity},\ }\href
  {https://doi.org/10.1023/A:1004894701900} {\bibfield  {journal} {\bibinfo
  {journal} {J. Stat. Phys.}\ }\textbf {\bibinfo {volume} {102}},\ \bibinfo
  {pages} {701} (\bibinfo {year} {2001})}\BibitemShut {NoStop}%
\bibitem [{\citenamefont {Fabricius}\ and\ \citenamefont
  {McCoy}(2001)}]{Fabricius2001}%
  \BibitemOpen
  \bibfield  {author} {\bibinfo {author} {\bibfnamefont {K.}~\bibnamefont
  {Fabricius}}\ and\ \bibinfo {author} {\bibfnamefont {B.~M.}\ \bibnamefont
  {McCoy}},\ }\bibfield  {title} {\bibinfo {title} {Bethe's equation is
  incomplete for the xxz model at roots of unity},\ }\href
  {https://doi.org/10.1023/A:1010380116927} {\bibfield  {journal} {\bibinfo
  {journal} {J. Stat. Phys.}\ }\textbf {\bibinfo {volume} {103}},\ \bibinfo
  {pages} {647} (\bibinfo {year} {2001})}\BibitemShut {NoStop}%
\bibitem [{\citenamefont {Popkov}\ \emph
  {et~al.}(2023{\natexlab{a}})\citenamefont {Popkov}, \citenamefont
  {Žnidarič},\ and\ \citenamefont {Zhang}}]{SHS_Dynamic}%
  \BibitemOpen
  \bibfield  {author} {\bibinfo {author} {\bibfnamefont {V.}~\bibnamefont
  {Popkov}}, \bibinfo {author} {\bibfnamefont {M.}~\bibnamefont {Žnidarič}},\
  and\ \bibinfo {author} {\bibfnamefont {X.}~\bibnamefont {Zhang}},\ }\bibfield
   {title} {\bibinfo {title} {{Universality in relaxation of spin helices under
  the $XXZ$- spin chain dynamics}},\ }\href
  {https://link.aps.org/doi/10.1103/PhysRevB.107.235408} {\bibfield  {journal}
  {\bibinfo  {journal} {Phys. Rev. B}\ }\textbf {\bibinfo {volume} {107}},\
  \bibinfo {pages} {235408} (\bibinfo {year} {2023}{\natexlab{a}})}\BibitemShut
  {NoStop}%
\bibitem [{\citenamefont {Braak}\ and\ \citenamefont
  {Andrei}(2001)}]{Braak2001}%
  \BibitemOpen
  \bibfield  {author} {\bibinfo {author} {\bibfnamefont {D.}~\bibnamefont
  {Braak}}\ and\ \bibinfo {author} {\bibfnamefont {N.}~\bibnamefont {Andrei}},\
  }\bibfield  {title} {\bibinfo {title} {On the spectrum of the xxz-chain at
  roots of unity},\ }\href {https://doi.org/10.1023/A:1012236111393} {\bibfield
   {journal} {\bibinfo  {journal} {J. Stats. Phys.}\ }\textbf {\bibinfo
  {volume} {105}},\ \bibinfo {pages} {677} (\bibinfo {year}
  {2001})}\BibitemShut {NoStop}%
\bibitem [{\citenamefont {Popkov}\ \emph
  {et~al.}(2023{\natexlab{b}})\citenamefont {Popkov}, \citenamefont {Zhang},\
  and\ \citenamefont {Klümper}}]{ChiralBasisXX}%
  \BibitemOpen
  \bibfield  {author} {\bibinfo {author} {\bibfnamefont {V.}~\bibnamefont
  {Popkov}}, \bibinfo {author} {\bibfnamefont {X.}~\bibnamefont {Zhang}},\ and\
  \bibinfo {author} {\bibfnamefont {A.}~\bibnamefont {Klümper}},\ }\bibfield
  {title} {\bibinfo {title} {Chiral bases for qubits and their applications to
  integrable spin chains},\ }\href {https://arxiv.org/abs/2303.14056}
  {\bibfield  {journal} {\bibinfo  {journal} {arXiv:2303.14056}\ } (\bibinfo
  {year} {2023}{\natexlab{b}})}\BibitemShut {NoStop}%
\bibitem [{\citenamefont {Schütz}(2023)}]{Schuetz2023}%
  \BibitemOpen
  \bibfield  {author} {\bibinfo {author} {\bibfnamefont {G.~M.}\ \bibnamefont
  {Schütz}},\ }\bibfield  {title} {\bibinfo {title} {A reverse duality for the
  asep with open boundaries},\ }\href
  {https://doi.org/10.1088/1751-8121/acda6a} {\bibfield  {journal} {\bibinfo
  {journal} {Journal of Physics A: Mathematical and Theoretical}\ }\textbf
  {\bibinfo {volume} {56}},\ \bibinfo {pages} {274001} (\bibinfo {year}
  {2023})}\BibitemShut {NoStop}%
\bibitem [{\citenamefont {Zhang}\ \emph
  {et~al.}(2015{\natexlab{a}})\citenamefont {Zhang}, \citenamefont {Li},
  \citenamefont {Cao}, \citenamefont {Yang}, \citenamefont {Shi},\ and\
  \citenamefont {Wang}}]{BetheStateXXZ}%
  \BibitemOpen
  \bibfield  {author} {\bibinfo {author} {\bibfnamefont {X.}~\bibnamefont
  {Zhang}}, \bibinfo {author} {\bibfnamefont {Y.-Y.}\ \bibnamefont {Li}},
  \bibinfo {author} {\bibfnamefont {J.}~\bibnamefont {Cao}}, \bibinfo {author}
  {\bibfnamefont {W.-L.}\ \bibnamefont {Yang}}, \bibinfo {author}
  {\bibfnamefont {K.}~\bibnamefont {Shi}},\ and\ \bibinfo {author}
  {\bibfnamefont {Y.}~\bibnamefont {Wang}},\ }\bibfield  {title} {\bibinfo
  {title} {{Bethe states of the XXZ spin-1/2 chain with arbitrary boundary
  fields}},\ }\href {https://doi.org/10.1016/j.nuclphysb.2015.01.022}
  {\bibfield  {journal} {\bibinfo  {journal} {Nucl. Phys. B}\ }\textbf
  {\bibinfo {volume} {893}},\ \bibinfo {pages} {70} (\bibinfo {year}
  {2015}{\natexlab{a}})}\BibitemShut {NoStop}%
\bibitem [{\citenamefont {Zhang}\ \emph
  {et~al.}(2015{\natexlab{b}})\citenamefont {Zhang}, \citenamefont {Li},
  \citenamefont {Cao}, \citenamefont {Yang}, \citenamefont {Shi},\ and\
  \citenamefont {Wang}}]{ZhangRetrieve}%
  \BibitemOpen
  \bibfield  {author} {\bibinfo {author} {\bibfnamefont {X.}~\bibnamefont
  {Zhang}}, \bibinfo {author} {\bibfnamefont {Y.-Y.}\ \bibnamefont {Li}},
  \bibinfo {author} {\bibfnamefont {J.}~\bibnamefont {Cao}}, \bibinfo {author}
  {\bibfnamefont {W.-L.}\ \bibnamefont {Yang}}, \bibinfo {author}
  {\bibfnamefont {K.}~\bibnamefont {Shi}},\ and\ \bibinfo {author}
  {\bibfnamefont {Y.}~\bibnamefont {Wang}},\ }\bibfield  {title} {\bibinfo
  {title} {{Retrieve the Bethe states of quantum integrable models solved via
  the off-diagonal Bethe Ansatz}},\ }\href
  {https://dx.doi.org/10.1088/1742-5468/2015/05/P05014} {\bibfield  {journal}
  {\bibinfo  {journal} {J. Stat. Mech.}\ }\textbf {\bibinfo {volume} {2015}},\
  \bibinfo {pages} {P05014} (\bibinfo {year} {2015}{\natexlab{b}})}\BibitemShut
  {NoStop}%
\end{thebibliography}

%

\setcounter{section}{0}
\renewcommand{\thesection}{\Alph{section}}
\setcounter{equation}{0}
\renewcommand{\theequation}{\Alph{section}.\arabic{equation}}
\renewcommand{\thesubsection}{\Alph{section}\arabic{subsection}}

\section{Elliptic theta functions}\label{Theta;Function}
We adopt the notations of elliptic theta functions $\vartheta_{\al}(u,q)$ from
Ref.~\cite{WatsonBook}
\begin{align}
\begin{aligned}
&\vartheta_{1}(u,q)=2\sum_{n=0}^\infty(-1)^n q^{(n+\frac12)^2}\sin[(2n+1)u],\\
&\vartheta_{2}(u,q)=2\sum_{n=0}^\infty q^{(n+\frac12)^2}\cos[(2n+1)u],\\
&\vartheta_{3}(u,q)=1+2\sum_{n=1}^\infty q^{n^2}\cos(2nu),\\
&\vartheta_{4}(u,q)=1+2\sum_{n=1}^\infty (-1)^nq^{n^2}\cos(2nu).
\end{aligned}
\end{align}
For convenience, we use the following shorthand notations $\ell{\al},\,\bell{\al}$
\begin{align}
\ell{\al}(u) \equiv  \vartheta_{\al} (\pi u,\eE^{\ir\pi\tau}),\quad \bell{\al}(u) \equiv   \vartheta_{\al} (\pi u ,\eE^{2\ir\pi\tau}),\quad {\rm Im}[\tau]>0,\quad \alpha=1,2,3,4,
\end{align}
The elliptic functions $\ell{\alpha}(u)$ possess important properties
\cite{WatsonBook,Wang-book}
\begin{align}
&\ell{2}(u)=\ell{1}(u+\tfrac12),\quad \ell{3}(u)=\eE^{\ir\pi(u+\frac{\tau}{4})}\ell{1}(u+\tfrac{1+\tau}{2}),\quad\ell{4}(u)=-\eE^{\ir\pi(u+\frac{\tau}{4}+\frac12)}\ell{1}(u+\tfrac{\tau}{2}), \\
&\ell{1}(-u)=-\ell{1}(u),\quad \ell{\alpha}(-u)=\ell{\alpha}(u),\quad \alpha=2,3,4,\\
&\ell{\alpha}(u+1)=-\ell{\alpha}(u),\quad \ell{\alpha'}(u+1)=\ell{\alpha'}(u),\quad \alpha=1,2,\quad \alpha'=3,4,\label{periodicity;1}\\
&\ell{\alpha}(u+\tau)=-\eE^{-\ir\pi(2u+\tau)}\ell{\alpha}(u),\quad \ell{\alpha'}(u+\tau)=\eE^{-\ir\pi(2u+\tau)}\ell{\alpha'}(u),\quad \alpha=1,4,\quad \alpha'=2,3.\label{periodicity;2}
\end{align}
Define the useful function
\begin{align}
\zeta(u)=\frac{\ell{1}'(u)}{\ell{1}(u)},\label{zeta}
\end{align}
which satisfies the following equations \cite{MPA2021,OpenXYZ2022}
\begin{align}
&\zeta(u)=-\zeta(-u),\qquad 
\zeta(u+1)=\zeta(u), \qquad \zeta(u+\tau)=\zeta(u)-2\ir\pi,\label{zeta;1}\\
&\frac{\ell{1}(v_1+v_2)\ell{1}(v_1+v_3)\ell{1}(v_2+v_3)}{\ell{1}(v_1)\ell{1}(v_2)\ell{1}(v_3)\ell{1}(v_1+v_2+v_3)}=\frac{1}{\ell{1}'(0)}\left[\zeta(v_1)+\zeta(v_2)+\zeta(v_3)-\zeta(v_1+v_2+v_3)\right].\label{zeta;sigma;4}
\end{align}
From Eq.~(\ref{zeta;sigma;4}), we get useful identities frequently used in
this paper
\begin{align}
&\frac{\ell{1}(v_1+v_2)\ell{1}(v_1+\eta)\ell{1}(v_2+\eta)}{\ell{1}(v_1)\ell{1}(v_2)\ell{1}(v_1+v_2+\eta)}=g(v_1)+g(v_2)+g(\eta)-g(v_1+v_2+\eta),\label{theta;zeta;1}\\
&\frac{\ell{2}(\eta)\ell{2}(u)\ell{1}(u\pm\eta)}{\ell{2}(0)\ell{1}(u)\ell{2}(u\pm\eta)}=\pm\left[g(u)\mp g(\eta)-g(u\pm\eta+\tfrac12)\right].\label{theta;zeta;2}
\end{align}

\section{Expressions of some functions}
\setcounter{equation}{0} Using the equations shown in Appendix
\ref{Theta;Function}, we derive the expression of some functions as follow
\begin{align}
a_+(u)&=-\left[g(u)+g(\eta)-g(u+\eta+\tfrac12)\right],\\
a_-(u)&= -\left[g(u)- g(\eta)-g(u-\eta+\tfrac12)\right],\\
A(u_{2\d+n})&=\sum_{k=1}^{n-1}w(u_{2\d+k})+\sum_{k=n+1}^Nw(u_{2\d-2+k})+w(-u_{2\d+n})-2a_-(u_{2\d+n})+2a_+(u_{2\d+n-1}) \no\\
&=E_0-2g(u_{2\d+n})+2g(u_{2\d+n-1})-2a_-(u_{2\d+n})+2a_+(u_{2\d+n-1})\no\\
&\overset{(\ref{theta;zeta;2})}{=} E_0+2\left[g(u_{2\d+n}+\tfrac12)-g(u_{2\d+n-1}+\tfrac12)-2g(\eta)\right],\\
B(\l,u)&=\frac{E(\l)-A(u)}{2}\no\\
&=g(\l-\tfrac{\eta}{2})+g(-\l-\tfrac{\eta}{2})+g(-u-\tfrac12)+g(u-\eta+\tfrac12)+2g(\eta)\no\\
&\overset{(\ref{theta;zeta;1})}{=}\frac{\ell{1}(\l+\frac{\eta}{2})}{\ell{1}(\l-\frac{\eta}{2})}\frac{\ell{2}(u-\eta)\ell{2}(\l-u-\tfrac{\eta}{2})}{\ell{2}(u)\ell{2}(\l-u+\tfrac{\eta}{2})}
+\frac{\ell{1}(\l-\frac{\eta}{2})}{\ell{1}(\l+\frac{\eta}{2})}\frac{\ell{2}(u)\ell{2}(\l-u+\tfrac{3\eta}{2})}{\ell{2}(u-\eta)\ell{2}(\l-u+\tfrac{\eta}{2})},\\
B(\l_1,\l_2,u,v)&=\frac{E(\l_1,\l_2)-A(u,v)}{2}\no\\
&=B(\l_1,u)+B(\l_2,v)=B(\l_1,v)+B(\l_2,u),\\
\overline{B}(\l_1,\l_2)&=\frac{E(\l_1,\l_2)-E_0+4g(\eta)}{2}\no\\
&=g(\l_1-\tfrac{\eta}{2})-g(\l_1+\tfrac{\eta}{2})+g(\l_2-\tfrac{\eta}{2})-g(\l_2+\tfrac{\eta}{2})+2g(\eta)\no\\
&\overset{(\ref{theta;zeta;1})}{=}\frac{\ell{1}(\l_1+\frac{\eta}{2})}{\ell{1}(\l_1-\frac{\eta}{2})}\frac{\ell{1}(\l_2+\frac{\eta}{2})}{\ell{1}(\l_2-\frac{\eta}{2})}\frac{\ell{1}(\l_1+\l_2-\eta)}{\ell{1}(\l_1+\l_2)}+\frac{\ell{1}(\l_1-\frac{\eta}{2})}{\ell{1}(\l_1+\frac{\eta}{2})}\frac{\ell{1}(\l_2-\frac{\eta}{2})}{\ell{1}(\l_2+\frac{\eta}{2})}\frac{\ell{1}(\l_1+\l_2+\eta)}{\ell{1}(\l_1+\l_2)}.
\end{align}

\section{Derivation of the XYZ eigenstates and BAE for the $M=2$ case} 
\label{app:M2Case}
\setcounter{equation}{0}

Acting Hamiltonian on the vectors $\{\ket{\d;n_1,n_2}\}$ we arrive at \small
\begin{align}
H\ket{\d;n_1,n_2}=&\,A(u_{2\d+n_1},u_{2\d+n_2-2})|\d;n_1,n_2\rangle+2A_-(u_{2\d+n_1})|\d;n_1-1,n_2\rangle+2A_+(u_{2\d+n_1})|\d;n_1+1,n_2\rangle\no\\
&+2A_-(u_{2\d+n_2-2})|\d;n_1,n_2-1\rangle+2A_+(u_{2\d+n_2-2})|\d;n_1,n_2+1\rangle,\quad 1<n_1<n_2<N,\quad n_2>n_1+1, \\
H\ket{\d;n,n+1}=&\,[E_0-4g(\eta)]|\d;n,n+1\rangle+2A_-(u_{2\d+n})|\d;n-1,n+1\rangle+2A_+(u_{2\d+n-1})|\d;n,n+2\rangle,\,\,\, 1<n<N-1,\\
H\ket{\d;n,N}=&\,A(u_{2\d+n},u_{2\d+N-2})|\d;n,N\rangle+2A_-(u_{2\d+n})|\d;n-1,N\rangle+2A_+(u_{2\d+n})|\d;n+1,N\rangle\no\\
&+2A_-(u_{2\d+N-2})|\d;n,N-1\rangle+2A_+(u_{2\d+2})|\d+1;1,n\rangle,\quad 1<n<N-1,\\
H\ket{\d;N-1,N}=&\,[E_0-4g(\eta)]|\d;N-1,N\rangle+2A_-(u_{2\d+N-1})|\d;N-2,N\rangle+2A_+(u_{2\d+2})|\d+1;1,N-1\rangle, \\
H\ket{\d;1,n}=&\,A(u_{2\d+1},u_{2\d+n-2})|\d;1,n\rangle+2A_-(u_{2\d+n-2})|\d;1,n-1\rangle+2A_+(u_{2\d+n-2})|\d;1,n+1\rangle\no\\&+2A_+(u_{2\d+1})|\d;2,n\rangle+2A_-(u_{2\d+N-3})\eE^{4\ir\pi L\eta}|\d-1;n,N\rangle,\quad 1<n<N-1,\\
H\ket{\d;1,2}=&\,[E_0-4g(\eta)]|\d;1,2\rangle+2A_+(u_{2\d})|\d;1,3\rangle+2A_-(u_{2\d+N-3})\eE^{4\ir\pi L\eta}|\d-1;2,N\rangle,\\
H\ket{\d;1,N}=&\,A(u_{2\d+1},u_{2\d+N-2})|\d;1,N\rangle+2A_-(u_{2\d+N-2})|\d;1,N-1\rangle+2A_+(u_{2\d+1})|\d;2,N\rangle,\\
H\ket{s;n,N}=&\,A(u_{2s+n},u_{2s+N-2})| s;n,N\rangle+2A_-(u_{2s+n})|s;n-1,N\rangle+2A_+(u_{2s+n})|s;n+1,N\rangle\no\\
&+2A_-(u_{2s+N-2})| s;n,N-1\rangle+2A_+(u_{0})\eE^{4\ir\pi L_0\eta}\widetilde{W}_n|0;1,n\rangle,\quad 1<n<N-1,\\
H\ket{s;N-1,N}=&\,[E_0-4g(\eta)]| s;N-1,N\rangle+2A_-(u_{2s+N-1})| s;N-2,N\rangle+2A_+(u_{0})\eE^{4\ir\pi L_0\eta}\widetilde{W}_{N-1}|0;1,N-1\rangle,\\
H\ket{0;1,n}=&\,A(u_{1},u_{n-2})|0;1,n\rangle+2A_-(u_{n-2})|0;1,n-1\rangle+2A_+(u_{n-2})|0;1,n+1\rangle\no\\
&+2A_+(u_{1})|0;2,n\rangle+2A_-(u_{2 s+N-1})\widetilde{W}_n^{-1}\eE^{4\ir\pi L\eta}| s;n,N\rangle,\quad 1<n<N-1,\\
H\ket{0;1,2}=&\,[E_0-4g(\eta)]|0;1,2\rangle+2A_+(u_{0})|\d;1,3\rangle+2A_-(u_{2s+N -1})\widetilde{W}_{2}^{-1}\eE^{4\ir\pi L\eta}| s;2,N\rangle,
\end{align}
\normalsize
where 
\begin{align}
A(u_1,u_2)=E_0+2\sum_{k=1}^2\left[g(u_k+\tfrac12)-g(u_k-\eta+\tfrac12)-2g(\eta)\right],
\end{align}
and 
\begin{align}
\widetilde{W}_n&=\exp\left\{-\ir\pi L_0\left[2N u_{s+1}+(N^2-7N+4n+4)\eta\right]\right\}.
\end{align}
The vectors in (\ref{Basis;M}) form a basis of the Hilbert space and the
corresponding eigenstates can be expanded as
\begin{align}
|\Psi(\l_1,\l_2)\rangle=
\sum_{\d=0}^{s}\,\sum_{1\leq n_1<n_2\leq N}F_{\d,n_1,n_2}(\l_1,\l_2)\ket{\d;n_1,n_2},\quad \mbox{with}\quad H|\Psi(\l_1,\l_2)\rangle=E(\l_1,\l_2)|\Psi(\l_1,\l_2)\rangle.
\end{align}
Here we assume
\begin{align}
E(\l_1,\l_2)=E_0+E_b(\l_1)+E_b(\l_2),
\end{align}
where $E_0$ and $E_b(\l)$ are defined in (\ref{E0}) and (\ref{Eb}) respectively.

For convenience, we assume that $F_{\d,n,n}\equiv F_{\d,n,n+N}\equiv 0$.  The
eigen equation of $H$ gives the following identities \small
\begin{align}
&F_{\d,n_1+1,n_2}(\l_1,\l_2)A_-(u_{2\d+n_1+1})+F_{\d,n_1-1,n_2}(\l_1,\l_2)A_+(u_{2\d+n_1-1})+F_{\d,n_1,n_2-1}(\l_1,\l_2)A_+(u_{2\d+n_2-3})\no\\
&+F_{\d,n_1,n_2+1}(\l_1,\l_2)A_-(u_{2\d+n_2-1})=B(\l_1,\l_2 ,u_{2\d+n_1},u_{2\d+n_2-2})F_{\d,n_1,n_2}(\l_1,\l_2),\quad 1<n_1<n_2<N,\quad n_2>n_1+1,\label{M2;recursive;1}\\
&F_{\d,n-1,n+1}(\l_1,\l_2)A_+(u_{2\d+n-1})+F_{\d,n,n+2}(\l_1,\l_2)A_-(u_{2\d+n})=\overline{B}(\l_1,\l_2)F_{\d,n,n+1}(\l_1,\l_2),\quad 1<n<N-1,\label{M2;recursive;2}\\
&F_{\d,n+1,N}(\l_1,\l_2)A_-(u_{2\d+n+1})+F_{\d,n-1,N}(\l_1,\l_2)A_+(u_{2\d+n-1})+F_{\d,n,N-1}(\l_1,\l_2)A_+(u_{2\d+N-3})\no\\
&+F_{\d+1,1,n}(\l_1,\l_2)\eE^{4\ir\pi L\eta}A_-(u_{2\d+N-1})=P(\l_1,\l_2 ,u_{2\d+n},u_{2\d+N-2})F_{\d,n,N}(\l_1,\l_2),\quad 1\leq n<N,\label{M2;recursive;3}\\
&F_{\d,2,n}(\l_1,\l_2)A_-(u_{2\d+2})+F_{\d-1,n,N}(\l_1,\l_2)A_+(u_{2\d})+F_{\d,1,n-1}(\l_1,\l_2)A_+(u_{2\d+n-3})\no\\
&+F_{\d,1,n+1}(\l_1,\l_2)A_-(u_{2\d+n-1})=P(\l_1,\l_2,u_{2\d+1},u_{2\d+n-2})F_{\d,1,n}(\l_1,\l_2),\quad 1<n\leq N,\label{M2;recursive;4},\\
&F_{s,n+1,N}(\l_1,\l_2)A_-(u_{2s+n+1})+F_{ s,n-1,N}(\l_1,\l_2)A_+(u_{2s+n-1})+F_{ s,n,N-1}(\l_1,\l_2)A_+(u_{2s+N-3})\no\\
&+F_{0,1,n}(\l_1,\l_2)\widetilde{W}_n^{-1}\eE^{4\ir\pi L\eta}A_-(u_{2s+N-1})=P(\l_1,\l_2 ,u_{2 s+n},u_{2s+N-2})F_{s,n,N}(\l_1,\l_2),\quad 1\leq n<N,\label{M2;recursive;5},\\
&F_{0,2,n}(\l_1,\l_2)A_-(u_{2})+F_{ s,n,N}(\l_1,\l_2)A_+(u_{0})\widetilde{W}_n\eE^{4\ir\pi L_0\eta}+F_{0,1,n-1}(\l_1,\l_2)A_+(u_{n-3})\no\\
&+F_{0,1,n+1}(\l_1,\l_2)A_-(u_{n-1})=P(\l_1,\l_2,u_{1},u_{n-2})F_{0,1,n}(\l_1,\l_2),\quad 1<n\leq N,\label{M2;recursive;6}
\end{align}
\normalsize
where 
\begin{align}
&B(\l_1,\l_2,u,v)=B(\l_1,u)+B(\l_2,v)=B(\l_1,v)+B(\l_2,u),\\
&\overline{B}(\l_1,\l_2)=\frac{\ell{1}(\l_1+\frac{\eta}{2})}{\ell{1}(\l_1-\frac{\eta}{2})}\frac{\ell{1}(\l_2+\frac{\eta}{2})}{\ell{1}(\l_2-\frac{\eta}{2})}\frac{\ell{1}(\l_1+\l_2-\eta)}{\ell{1}(\l_1+\l_2)}+\frac{\ell{1}(\l_1-\frac{\eta}{2})}{\ell{1}(\l_1+\frac{\eta}{2})}\frac{\ell{1}(\l_2-\frac{\eta}{2})}{\ell{1}(\l_2+\frac{\eta}{2})}\frac{\ell{1}(\l_1+\l_2+\eta)}{\ell{1}(\l_1+\l_2)},\\
&P(\l_1,\l_2,v+\eta,v)=\overline{B}(\l_1,\l_2),\quad P(\l_1,\l_2,u,v)=B(\l_1,\l_2,u,v),\quad u\neq v+\eta.
\end{align}

We propose the following ansatz 
\begin{align}
F_{\d,n_1,n_2}(\l_1,\l_2)=\alpha_\d\left[C_{1,2}\,U_{2\d+n_1}^{(n_1)}(\l_1)\,U_{2\d+n_2-2}^{(n_2)}(\l_2)+C_{2,1}\,U_{2\d+n_1}^{(n_1)}(\l_2)\,U_{2\d+n_2-2}^{(n_2)}(\l_1)\right],\label{M2;Ansatz}
\end{align}
where $U_{\d}^{(m)}(\l)$ is defined in (\ref{def;U}). 

Using the property of $U_{m}^{(n)}(\l)$ in Eq.~(\ref{U;property}), one can
prove that our ansatz (\ref{M2;Ansatz}) satisfies Eq.~(\ref{M2;recursive;1})
automatically. From Eq.~(\ref{M2;recursive;2}), we get
\begin{align}
&\left[C_{1,2}U_{k'+1}^{(n+1)}(\l_1)U_{k'-1}^{(n+1)}(\l_2)+C_{2,1}U_{k'+1}^{(n+1)}(\l_2)U_{k'-1}^{(n+1)}(\l_1)\right]A_-(u_{k'+1})\no\\
&+\left[C_{1,2}U_{k'}^{(n)}(\l_1)U_{k'-2}^{(n)}(\l_2)+C_{2,1}U_{k'}^{(n)}(\l_2)U_{k'-2}^{(n)}(\l_1)\right]A_+(u_{k'-2})\no\\
&=\left[B(\l_1,\l_2 ,u_{k'},u_{k'-1})-\overline{B}(\l_1,\l_2)\right]\left[C_{1,2}U_{k'}^{(n)}(\l_1)U_{k'-1}^{(n+1)}(\l_2)+C_{2,1}U_{k'}^{(n)}(\l_2)U_{k'-1}^{(n+1)}(\l_1)\right],\label{SM;Condition}
\end{align}
where
\begin{align}
&\left[B(\l_1,\l_2 ,u_{k},u_{k-1})-\overline{B}(\l_1,\l_2)\right]\no\\
&=2g(\eta)-g(u_{k}+\tfrac12)+g(u_{k-2}+\tfrac12)\no\\
&=\frac{\ell{2}(\eta)  \ell{2} (u_{k-1})\ell{1} (u_{k})}{\ell{2} (0) \ell{2} \left(u_{k}\right) \ell{1} (u_{k-1})}+\frac{\ell{2}(\eta) \ell{2} (u_{k-1}) \ell{1} (u_{k-2})}{\ell{2}(0) \ell{2} (u_{k-2}) \ell{1} (u_{k-1})}.
\end{align}
After tedious calculations, the ``two-body scattering matrix" $S$ is obtained,
see Appendix \ref{app:Smatrix}:
\begin{align}
S_{1,2}=\frac{C_{2,1}}{C_{1,2}}=\frac{\ell{1}(\l_1-\l_2-\eta)}{\ell{1}(\l_1-\l_2+\eta)}.\label{S;matrix}
\end{align}
To satisfy Eqs.~(\ref{M2;recursive;3}) - (\ref{M2;recursive;6}), the following
BAE for $\l_1,\,\l_2 $ and $\xi$ should be satisfied
\begin{align}
&\left[\frac{\ell{1}(\l_j+\frac{\eta}{2})}{\ell{1}(\l_j-\frac{\eta}{2})}\right]^N\prod_{k\neq j}\frac{\ell{1}(\l_j-\l_k-\eta)}{\ell{1}(\l_j-\l_k+\eta)}\exp\left(4\ir\pi L\l_j+\xi\right)=1,\quad j=1,2,\label{BAE;M2;1}\\
&\exp[4\ir\pi L_0(\l_1+\l_2)-(s+1)\xi]=1,\label{BAE;M2;2}
\end{align} 
The coefficient $\alpha_\d$ in (\ref{M2;Ansatz}) is parameterized in terms of
$\xi$ and $\d$ as
\begin{align}
\alpha_\d=\exp\left[2\ir\pi L \d\left(2u_{\d}+2L\tau+\eta\right)- \d\xi\right].
\end{align}

\section{Derivation of the $S$-matrix in (\ref{S;matrix})}
\label{app:Smatrix}
\setcounter{equation}{0}
Introduce the useful identity \cite{Wang-book}
\begin{align}
&\ell{1}(u+x)\ell{1}(u-x)\ell{1}(v+y)\ell{1}(v-y)-\ell{1}(u+y)\ell{1}(u-y)\ell{1}(v+x)\ell{1}(v-x)\no\\
&=\ell{1}(u+v)\ell{1}(u-v)\ell{1}(x+y)\ell{1}(x-y).\label{Riemann}
\end{align}
Eq.~(\ref{SM;Condition}) can be rewritten as 
\begin{align}
	J(\l_1,\l_2) C_{1,2}+J(\l_2,\l_1)C_{2,1}=0.
\end{align}
The expression of $J(\l_1,\l_2)$ can be derived as follow
\begin{align}
J(\l_1,\l_2)&=A_-(u_{k'+1}){U_{k'+1}^{(n+1)}(\l_1)U_{k'-1}^{(n+1)}(\l_2)}+A_+(u_{k'-2}){U_{k'}^{(n)}(\l_1)U_{k'-2}^{(n)}(\l_2)}\no\\
&\quad-\left[B(\l_1,\l_2 ,u_{k'},u_{k'-1})-\overline{B}(\l_1,\l_2)\right]{U_{k'}^{(n)}(\l_1)U_{k'-1}^{(n+1)}(\l_2)}\no\\
&=\left[\frac{\ell{1}(\l_1+\frac{\eta}{2})}{\ell{1}(\l_1-\frac{\eta}{2})}\frac{\ell{1}(\l_2+\frac{\eta}{2})}{\ell{1}(\l_2-\frac{\eta}{2})}\right]^{n+1}\frac{\ell{2}(\l_1-u_{k'+1}+\frac{\eta}{2})\ell{2}(\l_2-u_{k'-1}+\frac{\eta}{2})}{\ell{2}^2(u_{k'})\ell{2}(u_{k'-2})\ell{2}(u_{k'-1})}\no\\
&\quad+\left[\frac{\ell{1}(\l_1+\frac{\eta}{2})}{\ell{1}(\l_1-\frac{\eta}{2})}\frac{\ell{1}(\l_2+\frac{\eta}{2})}{\ell{1}(\l_2-\frac{\eta}{2})}\right]^{n}\frac{\ell{2}(\l_1-u_{k'}+\frac{\eta}{2})\ell{2}(\l_2-u_{k'-2}+\frac{\eta}{2})}{\ell{2}(u_{k'})\ell{2}(u_{k'-1})\ell{2}^2(u_{k'-2})}\no\\
&\quad-\left[\frac{\ell{1}(\l_1+\frac{\eta}{2})}{\ell{1}(\l_1-\frac{\eta}{2})}\right]^{n}\left[\frac{\ell{1}(\l_2+\frac{\eta}{2})}{\ell{1}(\l_2-\frac{\eta}{2})}\right]^{n+1}\frac{\ell{2}(\l_1-u_{k'}+\frac{\eta}{2})\ell{2}(\l_2-u_{k'-1}+\frac{\eta}{2})}{\ell{2}^2(u_{k'})\ell{2}(u_{k'-1})\ell{2}(u_{k'-2})}\frac{\ell{2}(\eta)  \ell{1} (u_{k'})}{\ell{2} (0)  \ell{1} (u_{k'-1})}\no\\
&\quad-\left[\frac{\ell{1}(\l_1+\frac{\eta}{2})}{\ell{1}(\l_1-\frac{\eta}{2})}\right]^{n}\left[\frac{\ell{1}(\l_2+\frac{\eta}{2})}{\ell{1}(\l_2-\frac{\eta}{2})}\right]^{n+1}\frac{\ell{2}(\l_1-u_{k'}+\frac{\eta}{2})\ell{2}(\l_2-u_{k'-1}+\frac{\eta}{2})}{\ell{2}(u_{k'})\ell{2}(u_{k'-1})\ell{2}^2(u_{k'-2})}\frac{\ell{2}(\eta) \ell{1} (u_{k'-2})}{\ell{2}(0)  \ell{1} (u_{k'-1})}\no\\
&=\chi\,\frac{\ell{1}(\l_2-u_{k'-1}+\tfrac{1+\eta}{2})}{\ell{1}(u_{k'}+\frac12)}\frac{\ell{1}(\l_2+\frac{\eta}{2})}{\ell{1}(\l_2-\frac{\eta}{2})}\left\{\frac{\ell{1}(\l_1+\frac{\eta}{2})}{\ell{1}(\l_1-\frac{\eta}{2})}\ell{1}(\l_1-u_{k'+1}+\tfrac{1+\eta}{2})\right.\no\\
&\quad\left.-\frac{\ell{1}(\eta+\frac12) \ell{1} (u_{k'})}{\ell{1} (\frac12)  \ell{1} (u_{k'-1})}\ell{1}(\l_1-u_{k'}+\tfrac{1+\eta}{2})\right\}
+\chi\,\frac{\ell{1}(\l_1-u_{k'}+\frac{1+\eta}{2})}{\ell{1}(u_{k'-2}+\frac12)}\no\\
&\quad\times\left[\ell{1}(\l_2-u_{k'-2}+\tfrac{1+\eta}{2})-\frac{\ell{1}(\l_2+\frac{\eta}{2})}{\ell{1}(\l_2-\frac{\eta}{2})}\frac{\ell{1}(\eta+\frac12)\ell{1}(u_{k'-2})}{\ell{1}(\frac12)\ell{1}(u_{k'-1})}\ell{1}(\l_2-u_{k'-1}+\tfrac{1+\eta}{2})\right]\no\\
&\overset{(\ref{Riemann})}{=}-\chi\,\frac{\ell{1}(\eta)\ell{1}(\l_2+\frac{\eta}{2})\ell{1}(\l_2-u_{k'-1}+\tfrac{1+\eta}{2})\ell{1}(\l_1-u_{k'}+\tfrac{\eta}{2})\ell{1}(\l_1-\tfrac{\eta}{2}+\tfrac{1}{2})}{\ell{1}(\l_2-\frac{\eta}{2})\ell{1}(\l_1-\frac{\eta}{2})\ell{1}(\frac12)\ell{1}(u_{k'-1})}\frac{}{}\no\\
&\quad+\chi\,\frac{\ell{1}(\eta)\ell{1}(\l_1-u_{k'}+\frac{1+\eta}{2})\ell{1}(\l_2+\frac{\eta}{2}+\frac12)\ell{1}(\l_2-u_{k'}+\frac{3\eta}{2})}{\ell{1}(\l_2-\frac{\eta}{2})\ell{1}(\frac12)\ell{1}(u_{k'-1})}\no\\
&\overset{(\ref{Riemann})}{=}\chi\,\frac{\ell{1}(\eta)\ell{1}(\l_2-\l_1+\eta)\ell{2}(\l_1+\l_2-u_{k'-1})}{\ell{1}(\l_2-\frac{\eta}{2})\ell{1}(\l_1-\frac{\eta}{2})},
\end{align}
where 
\begin{align}
\chi=\left[\frac{\ell{1}(\l_1+\frac{\eta}{2})}{\ell{1}(\l_1-\frac{\eta}{2})}\frac{\ell{1}(\l_2+\frac{\eta}{2})}{\ell{1}(\l_2-\frac{\eta}{2})}\right]^{n}\ell{2}^{-1}(u_{k'})\ell{2}^{-1}(u_{k'-1})\ell{2}^{-1}(u_{k'-2}).\label{chi}
\end{align}
Thus, we get the two-body scattering matrix 
\begin{align}
S_{1,2}=\frac{C_{2,1}}{C_{1,2}}=-\frac{J(\l_1,\l_2)}{J(\l_2,\l_1)}=\frac{\ell{1}(\l_1-\l_2-\eta)}{\ell{1}(\l_1-\l_2+\eta)}.
\end{align}

\section{Derivation of BAE}\setcounter{equation}{0}

The quasi-periodicity of $\ell{\alpha}(u)$ gives rise to
\begin{align}
\frac{\ell{\alpha}(u+2l\tau+2k)}{\ell{\alpha}(u)}
=\exp\left[-4\ir\pi l(u+l\tau)\right],\quad \alpha=1,2.
\end{align}
When \begin{align}
(N-2M)\eta=2L\tau+2K,\qquad 2(s+1)\eta=2L_0\tau+2K_0,
\end{align}
it is straightforward to get 
\begin{align}
&\frac{U_{m+2s+2}^{(n)}(\l)}{U_{m}^{(n)}(\l)}
=\exp\left[2\ir\pi L_0(2\l+2u_{m+s+1}-\eta)\right],\label{U;period;1}\\
&\frac{U_{m+N-2M}^{(n+N)}(\l)}{U_{m}^{(n)}(\l)}
=\left[\frac{\ell{1}(\l+\frac{\eta}{2})}{\ell{1}(\l-\frac{\eta}{2})}\right]^N\exp\left[2\ir\pi L(2\l+2u_{m}+2L\tau-\eta)\right].\label{U;period;2}
\end{align}

\subsection{$M=1$ case}
Property (\ref{U;property}) of $U_m^{(n)}(\l)$ ensures that the first of
Eq.~(\ref{recursive;3}) is satisfied for arbitrary $m,n$.  The remaining
(\ref{recursive;3}) give the following identities:
\begin{align}
&\eE^{-4\ir\pi L\eta}\frac{F_{\d,N+1}(\l)}{F_{\d+1,1}(\l)}=\frac{F_{\d,N}(\l)}{F_{\d+1,0}(\l)}=1,\label{M1;Eq1}\\
&\eE^{-4\ir\pi L_0\eta}\frac{F_{0,0}(\l)}{F_{s+1,0}(\l)}=\frac{F_{0,1}(\l)}{F_{s+1,1}(\l)}=W.\label{M1;Eq2}
\end{align}
The BAE (\ref{BAE;M1;1}), (\ref{BAE;M1;2}) are derived from the above
equations directly.

\subsection{$M=2$ case}
With the help of Eqs.~(\ref{U;property}) and (\ref{S;matrix}), functional
relations (\ref{M2;recursive;1}) and (\ref{M2;recursive;2}) hold for arbitrary
$\d$, $n_1,n_2$. Thus, to satisfy Eqs.~(\ref{M2;recursive;3}) -
(\ref{M2;recursive;6}), one has
\begin{align}
&\frac{F_{\d,n,N}(\l_1,\l_2)}{F_{\d+1,0,n}(\l_1,\l_2)}=\eE^{-4\ir\pi L\eta}\frac{F_{\d,n,N+1}(\l_1,\l_2)}{F_{\d+1,1,n}(\l_1,\l_2)}=1,\\
&\frac{F_{s+1,0,n}(\l_1,\l_2)}{F_{0,0,n}(\l_1,\l_2)}=\eE^{-4\ir\pi L_0\eta}\frac{F_{s+1,1,n}(\l_1,\l_2)}{F_{0,1,n}(\l_1,\l_2)}=\widetilde{W}^{-1}_n\exp(-4\ir\pi L_0\eta).
\end{align}
The above equations are equivalent to
\begin{align}
&\frac{\alpha_\d\,C_{1,2}\,U_{2\d+N-2}^{(N)}(\l_2)}{\alpha_{\d+1}\,C_{2,1}\,U_{2\d+2}^{(0)}(\l_2)}=\frac{\alpha_\d\,C_{2,1}\,U_{2\d+N-2}^{(N)}(\l_1)}{\alpha_{\d+1}\,C_{1,2}\,U_{2\d+2}^{(0)}(\l_1)}=1,\\
&\frac{\alpha_{s+1}\,U_{2s+n}^{(n)}(\l_1)\,U_{2s+2}^{(0)}(\l_2)}{\alpha_0\,U_{n-2}^{(n)}(\l_1)\,U_{0}^{(0)}(\l_2)}=\frac{\alpha_{s+1}\,U_{2s+n}^{(n)}(\l_2)\,U_{2s+2}^{(0)}(\l_1)}{\alpha_0\,U_{n-2}^{(n)}(\l_2)\,U_{0}^{(0)}(\l_1)}=\widetilde{W}^{-1}_n\exp(-4\ir\pi L_0\eta).
\end{align}
Using Eqs.~(\ref{U;period;1}), (\ref{U;period;2}), we finally arrive at the
BAE (\ref{BAE;M2;1}), (\ref{BAE;M2;2}) after straightforward calculations.

\section{Degeneration of the inhomogeneous BAE}\label{AppendixE}\setcounter{equation}{0}
For the periodic XYZ chain, the eigenvalue of the transfer matrix $\Lambda(u)$ can be
given by the following inhomogeneous $T$-$Q$ relation
\cite{Wang-book,Cao2013off}
\begin{align}
\Lambda(u)&=\eE^{2\ir\pi l_0u+\kappa}\,\ell{1}^N(u+\eta)\frac{\mathcal{Q}_1(u-\eta)}{\mathcal{Q}_2(u)}+\eE^{-2\ir\pi l_0(u+\eta)-\kappa}\,\ell{1}^N(u)\frac{\mathcal{Q}_2(u+\eta)}{\mathcal{Q}_1(u)}\no\\
&\quad +c\,\ell{1}^m(u+\tfrac{\eta}{2})\frac{\ell{1}^N(u+\eta)\ell{1}^N(u)}{\mathcal{Q}_1(u)\mathcal{Q}_2(u)},\label{TQ}
\end{align}
where $l_0$ is an integer, and $M_0$ and $m$ are two non-negative integers
which satisfy the relation $N+2m=2M_0$.  The $\mathcal Q$-functions
$\mathcal{Q}_{1,2}(u)$ are defined by
\begin{align}
\mathcal{Q}_1(u)=\prod_{j=1}^{M_0}\ell{1}(u-\mu_j),\quad \mathcal{Q}_2(u)=\prod_{j=1}^{M_0}\ell{1}(u-\nu_j).
\end{align}
The BAE for the generic periodic XYZ chain read \cite{Wang-book,Cao2013off} 
\begin{align}
&\exp\left\{\ir\pi\left[\left(N-2M_0\right)\eta-2\sum_{j=1}^{M_0}(\mu_j-\nu_j)\right]\right\}=\exp(2\ir\pi l_0\tau),\\
&c\,\exp\left\{2\ir\pi\left[M_0\eta+\sum_{j=1}^{M_0}(\mu_j+\nu_j)\right]\right\}=c,\\
&\frac{c\,\ell{1}^m(\mu_j+\frac{\eta}{2})\ell{1}^N(\mu_j+\eta)}{\exp[-2\ir\pi l_0(\mu_j+\eta)-\kappa]}=-\prod_{l=1}^{M_0}\ell{1}(\mu_j-\nu_l+\eta)\ell{1}(\mu_j-\nu_l),\\
&\frac{c\,\ell{1}^m(\nu_j+\frac{\eta}{2})\ell{1}^N(\nu_j)}{\exp(2\ir\pi l_0\nu_j+\kappa)}=-\prod_{l=1}^{M_0}\ell{1}(\nu_j-\mu_l-\eta)\ell{1}(\nu_j-\mu_l),\\
&\eE^{\kappa}\prod_{j=1}^{M_0}\frac{\ell{1}(\mu_j+\eta)}{\ell{1}(\nu_j)}=\eE^{\frac{2\ir\pi k}{N}},\quad k\in\mathbb{Z}.
\end{align}

Under the condition
\begin{align}
(N-2M)\eta=2L\tau+2K,\quad L, K\in \mathbb{Z}, \label{ABA;Degeneration}
\end{align}
the Bethe roots have to satisfy the following relations 
\begin{align}
&\mu_j=\nu_j\equiv \l_j-\frac{\eta}{2},\quad j=1,\ldots,M,\\
&\mu_{j+M}=\nu_{j+M}-\eta,\quad j=1,\ldots,M_0-M.
\end{align}
The inhomogeneous term vanishes with $l_0=L$ and the resulting homogeneous BAE
are
\begin{align}
&\left[\frac{\ell{1}(\l_j+\frac{\eta}{2})}{\ell{1}(\l_j-\frac{\eta}{2})}\right]^N\prod_{k\neq j}^M\frac{\ell{1}(\l_j-\l_k-\eta)}{\ell{1}(\l_j-\l_k+\eta)}\exp\left(4\ir\pi L\l_j+2\kappa\right)=1,\quad j=1,2,\ldots,M,\label{ABAE;1}\\
&\eE^{\kappa}\prod_{j=1}^{M}\frac{\ell{1}(\l_j+\frac{\eta}{2})}{\ell{1}(\l_j-\frac{\eta}{2})}=\eE^{\frac{2\ir\pi k}{N}},\qquad k\in\mathbb{Z}.\label{ABAE;2}
\end{align} 
From Eqs.~(\ref{ABAE;1}) - (\ref{ABAE;2}) we get 
\begin{align}
&\prod_{j=1}^M\left[\frac{\ell{1}(\l_j+\frac{\eta}{2})}{\ell{1}(\l_j-\frac{\eta}{2})}\right]^N\exp\left(4\ir\pi L\sum_{j=1}^M\l_j+2M\kappa\right)=1,\no\\
&\eE^{N\kappa}\prod_{j=1}^{M}\left[\frac{\ell{1}(\l_j+\frac{\eta}{2})}{\ell{1}(\l_j-\frac{\eta}{2})}\right]^N=1,
\end{align}
leading to 
\begin{align}
\exp\left[4\ir\pi L\sum_{j=1}^M\l_j-(N-2M)\kappa\right]=1.\label{ABAE;3}
\end{align}
The case we consider in this paper belongs to the degenerate case
(\ref{ABA;Degeneration}) and an additional identity
(\ref{Constraint;Periodic;2}) is needed.  By letting
$\eE^{\kappa}=\pm\eE^{\frac{\xi}{2}}$, we see that our BAE in
(\ref{BAE;M;1}) - (\ref{BAE;M;2}) are consistent with Eqs.~(\ref{ABAE;1}) and
(\ref{ABAE;3}).

Under the condition (\ref{ABA;Degeneration}), the $T$-$Q$ relation (\ref{TQ})
reduces to a conventional homogeneous one
\begin{align}
\Lambda(u)&=\eE^{2\ir\pi Lu+\kappa}\,\ell{1}^N(u+\eta)\prod_{j=1}^M\frac{\ell{1}(u-\l_j-\frac{\eta}{2})}{\ell{1}(u-\l_j+\frac{\eta}{2})}+\eE^{-2\ir\pi L(u+\eta)-\kappa}\,\ell{1}^N(u)\prod_{j=1}^M\frac{\ell{1}(u-\l_j+\frac{3\eta}{2})}{\ell{1}(u-\l_j+\frac{\eta}{2})},\label{TQ3}
\end{align}
where $\{\l_j\}$ and $\kappa$ are given by Eqs.~(\ref{ABAE;1}),
(\ref{ABAE;3}).  \\

\textit{Remark:} When $M=0$ and $N\eta=2L\tau+2K,\,\,L, K\in \mathbb{Z}$,
$\Lambda(u)$ in (\ref{TQ3}) reads
\begin{align}
\Lambda(u)&=\eE^{2\ir\pi Lu+\kappa}\,\ell{1}^N(u+\eta)+\eE^{-2\ir\pi L(u+ \eta)-\kappa}\,\ell{1}^N(u),\quad \kappa=\tfrac{2\ir\pi k}{N},\quad k\in\mathbb{Z}.\label{TQ2}
\end{align}  
The selection of $\kappa$ in (\ref{TQ2}) shows that the transfer matrix has
$N$ factorized eigenstates which are exactly our elliptic spin-helix states in
(\ref{M0;eigenstate}) \cite{Baxter-book,Takhtadzhan1979}, and they all
correspond to the same energy $E_0$ in (\ref{E0}).  Replacing $\eta,L,K$
with $-\eta,-L,-K$, leaves the Hamiltonian invariant. Therefore, we can
construct another set of $N$ independent elliptic spin-helix states with different
chirality.  Consequently, the minimal degeneracy of $E_0$ is $2N$.

\section{Baxter's generalized Bethe ansatz}\label{Basis;Baxter}
\label{app:Baxter}
\setcounter{equation}{0} Baxter proposed a basis and a Bethe ansatz solution
for the periodic XYZ chain in \cite{Baxter6,Baxter4}. In this section, we will
recall his results and then explain the relation between our Bethe ansatz and
Baxter's.

First, recall Baxter's result in \cite{Baxter6}. Define a local state
\begin{align} 
\psi'(u)=\binom{\bell{1}(u)}{\bell{4}(u)}.\label{Psi2}
\end{align}
Let us introduce the following notations
\begin{align}
&r_m=r-v-\tfrac{1}{2}\eta+m\eta,\quad t_m=t+v-\tfrac{3}{2}\eta+m\eta,\quad\omega_m=\frac{r+t}{2}+m\eta,\label{r;t}
\end{align} 
where $r$, $t$ and $v$ are arbitrary.  Then the following global state is
constructed
\begin{align}
\ket{\d;n_1,\ldots,n_M}_{I}&=\bigotimes_{k_1=1}^{n_1-1}\psi'(r_{d+k_1})\bigotimes \psi'(t_{d+n_1})\bigotimes_{k_2=n_1+1}^{n_2-1}\psi'(r_{d+k_2-2})\bigotimes \psi'(t_{\d+n_2-2})\cdots\no\\
&\quad \bigotimes \psi'(t_{\d+n_M-2M+2})\bigotimes_{k_{M+1}=n_M+1}^{N}\psi'(r_{\d+k_{M+1}-2M}).\label{BaxterBasis1}
\end{align} 
Under the condition 
\begin{align}
Q\eta=2m_1+2m_2\tau,\quad (N-2M')\eta=2m'_1+2m'_2\tau,\quad Q\in\mathbb{N^+},\quad m_1,m_2,m'_1,m'_2\in\mathbb{Z},\label{Constraint;Baxter}
\end{align} 
Baxter has proved that the following states 
\begin{align}
\ket{\d;n_1,n_2,\ldots,n_{M'}}_I,\quad\d=1,2,\ldots,Q,\quad 1\leq n_1< n_2<\ldots<n_{M'}\leq N,\label{BaxterBasis2}
\end{align} 
form a closed subspace for the transfer matrix of the eight-vertex model,
corresponding to the XYZ Hamiltonian \cite{Baxter6}.

When $M'=\frac{N}{2}$, the eigenstate of the Hamiltonian can be expanded as
\cite{Baxter6,Baxter4}
\begin{align}
&\ket{\Psi}_I=\sum_{d=1}^Q\sum_{\substack{1\leq n_1<n_2<\cdots\\\cdots<n_{M'}\leq N}}\sum_P \varpi^d\,Y(P)\, G_{p_1}(\d,n_1)G_{p_2}(\d-2,n_2)\cdots \no\\[2pt]
&\qquad \qquad \cdots G_{p_{M'}}(\d-2M'+2,n_{M'}) \ket{\d;n_1,n_2,\ldots,n_{M'}}_I,\label{Baxter;BS1}\\[2pt]
&G_{j}(\d,x)=\left[\frac{\ell{1}(\mu_j+\frac\eta 2)}{\ell{1}(\mu_j-\frac\eta 2)}\right]^x\frac{\ell{2}(\mu_j-\omega_{l+x-1}+\frac{\eta}{2})}{\ell{2}(\omega_{l+x-1})\ell{2}(\omega_{l+x-2})},\label{G;function}\\[2pt]
&Y(P)=\epsilon_P\prod_{1\leq j\leq m\leq M'}\ell{1}(\mu_{p_j}-\mu_{p_m}+\eta),\label{Y;function}
\end{align}
where $P=\{p_1,\ldots,p_{M'}\}$ is the permutation of integers
$\{1,\ldots,M'\}$, $\epsilon_P$ is the signature of the permutation $P$ and
$\omega_{m}$ is defined in (\ref{r;t}). The Bethe roots
$\{\mu_1,\ldots,\mu_{M'}\}$ satisfy the BAE
\begin{align}
\varpi^{-2}\,\left[\frac{\ell{1}(\mu_j+\frac{\eta}{2})}{\ell{1}(\mu_j-\frac{\eta}{2})}\right]^N\prod_{k\neq j}^{N/2}\frac{\ell{1}(\mu_j-\mu_k-\eta)}{\ell{1}(\mu_j-\mu_k+\eta)}=1,\quad \varpi^Q=1,\quad  j=1,\ldots,N/2.\label{Baxter;BAE}
\end{align}
Our Bethe ansatz equations for the $M=\frac{N}{2}$ case are 
\begin{align}
&\eE^{\xi}\left[\frac{\ell{1}(\l_j+\frac{\eta}{2})}{\ell{1}(\l_j-\frac{\eta}{2})}\right]^N\,\prod_{k\neq j}^{N/2}\frac{\ell{1}(\l_j-\l_k-\eta)}{\ell{1}(\l_j-\l_k+\eta)}=1,\quad j=1,2,\ldots,N/2,\label{BAE;M;3}\\
&\exp\left(4\ir\pi L_0\sum_{k=1}^{N/2}\l_k-(s+1)\xi\right)=1.\label{BAE;M;4}
\end{align}
\textit{Remark:} When  $\eta=\frac{m}{m'}(L_0=0),\,\,
m,m'\in\mathbb{Z}$, we find that the solutions of the BAE (\ref{Baxter;BAE})
and (\ref{BAE;M;3}), (\ref{BAE;M;4}) have the following correspondence
\begin{align}
\{\mu_1,\ldots,\mu_{N/2}\}=\{\l_1,\ldots,\l_{N/2}\},\quad Q=2(s+1),\quad \varpi=\pm \eE^{-\frac{\xi}{2}}.\label{Correspondence}
\end{align} 
For a nonzero $L_0$ in Eq. (\ref{BAE;M;4}), i.e.~$\rm{Im}[\eta]\neq 0$,
  our BAE appear to be \textit{not} equivalent to Baxter's. The parameter $\varpi$ in
  Baxter's BAE (\ref{Baxter;BAE}) is always a root of unity while the
  parameter $\eE^{-\frac{\xi}{2}}$ in our BAE (\ref{BAE;M;3}), (\ref{BAE;M;4})
  is not for a nonzero $L_0$. Obvious examples are the right panels of
  Tab. \ref{N2;data} and Tab. \ref{N4;data}. This inconsistency seems to contradict
  Eq. (\ref{Correspondence}): for $M=\frac{N}{2}=2$ and $L_0\neq 0$
  case, we prove numerically that the solution of our BAE (\ref{BAE;M;3}),
  (\ref{BAE;M;4}) and the correspondence Eq.~(\ref{Correspondence}) still give
  the correct expansion coefficients in (\ref{Baxter;BS1}). The apparent
  disagreement can be understood.

 By applying the conjugate modulus transformation we find that the same XYZ
 chain with parameters $\eta, \tau$ can be parameterized with $\eta/\tau,
 -1/\tau$ plus a unitary transformation interchanging the $J_x$ and $J_z$
 coefficients as following 
\begin{align}
&\oell{\alpha}(u)= \vartheta_{\al} (\pi u,\eE^{-\frac{\ir\pi}{\tau}}),\quad \alpha=1,2,3,4,\no\\
&\ell{1}(u)=\ir\sqrt{\tfrac{\ir}{\tau}}\,\eE^{-\frac{\ir\pi u^2}{\tau}}\oell{1}(\tfrac{u}{\tau}),\quad \ell{\alpha}(u)=\sqrt{\tfrac{\ir}{\tau}}\,\eE^{-\frac{\ir\pi u^2}{\tau}}\oell{6-\alpha}(\tfrac{u}{\tau}),\quad \alpha=2,3,4,\\
&J_x|_{\eta,\tau}=\eE^{-\frac{\ir\pi \eta^2}{\tau}}J_z|_{\eta/\tau,-1/\tau},\quad J_y|_{\eta,\tau}=\eE^{-\frac{\ir\pi \eta^2}{\tau}}J_y|_{\eta/\tau,-1/\tau},\quad J_z|_{\eta,\tau}=\eE^{-\frac{\ir\pi \eta^2}{\tau}}J_x|_{\eta/\tau,-1/\tau}.
\end{align} 
Interestingly, $\tau$ and $-1/\tau$ are from $\ir{\mathbb R}^+$
 and for real (imaginary) value of $\eta$ the parameter $\eta/\tau$ is
 imaginary (real). This means that the case of imaginary $\eta$ is transformed
 into the case of Baxter's treatment where he uses real values for this
 parameter. Now the question comes up how the Bethe ansatz equations
 transform? When $\eta$ is purely imaginary, we do the conjugate modulus transformation for the BAE (\ref{BAE;M;3})
\begin{align}
&\exp\left(\xi-\frac{4\ir\pi\eta}{\tau}\sum_{k=1}^{N/2}\l_k\right)\left[\frac{\oell{1}(\frac{1}{\tau}(\l_j+\frac{\eta}{2}))}{\oell{1}(\frac{1}{\tau}(\l_j-\frac{\eta}{2}))}\right]^N\,\prod_{k\neq j}^{N/2}\frac{\oell{1}(\frac{1}{\tau}(\l_j-\l_k-\eta))}{\oell{1}(\frac{1}{\tau}(\l_j-\l_k+\eta))}\no\\
&\overset{(\ref{BAE;M;4})}{=}\exp\left(\tfrac{2\ir\pi n}{s+1}\right)\left[\frac{\oell{1}(\frac{1}{\tau}(\l_j+\frac{\eta}{2}))}{\oell{1}(\frac{1}{\tau}(\l_j-\frac{\eta}{2}))}\right]^N\,\prod_{k\neq j}^{N/2}\frac{\oell{1}(\frac{1}{\tau}(\l_j-\l_k-\eta))}{\oell{1}(\frac{1}{\tau}(\l_j-\l_k+\eta))}=1,\quad j=1,2,\ldots,N/2,\quad n\in\mathbb{Z}.
\end{align}
The real part of the parameter $\xi$ in (\ref{BAE;M;4}) now disappears and a root
 of unity factor results.

\bigskip

By letting $Q$ in (\ref{Constraint;Baxter}) be $2(s+1)$, we can divide the
states in (\ref{BaxterBasis2}) into two subsets
\begin{align}
&\left\{\ket{\d;n_1,n_2,\ldots,n_{M'}}_I\right\} =\left\{\ket{\d';n_1,n_2,\ldots,n_{M'}}_{I\!I}\right\}\cup\left\{\ket{\d';n_1,n_2,\ldots,n_{M'}}_{I\!I\!I}\right\},\quad \d=1,\ldots,Q,\,\, d'=1,\ldots,(s+1),\no\\[2pt]
&\ket{\d';n_1,n_2,\ldots,n_{M'}}_{I\!I}=\ket{2\d'-1;n_1,n_2,\ldots,n_{M'}}_{I},\quad\ket{\d';n_1,n_2,\ldots,n_{M'}}_{I\!I\!I}=\ket{2\d';n_1,n_2,\ldots,n_{M'}}_{I},\label{TwoSets}
\end{align}
One sees that the state $\ket{\d;n_1,n_2,\ldots,n_{M'}}_{I\!I\!I}$ can be
obtained from $\ket{\d;n_1,n_2,\ldots,n_{M'}}_{I\!I}$ by shifting an overall
phase for all local states as $\psi'(u)\to\psi'(u+\eta)$.  Since $\psi'(u)$
is a linear combination of $\psi'(v)$ and $\psi'(v+2\eta)$ with
$u,v\in\mathbb{C}$ as
\begin{align}
&\psi'(u)=\beta_1(u,v)\psi'(v+2\eta)+\beta_2(u,v)\psi'(v),\\
&\beta_1(u,v)=\frac{\ell{1}(\frac{u-v}{2}) \ell{2}(\frac{u+v}{2})}{\ell{1}(\eta) \ell{2}(v+\eta)},\quad \beta_2(u,v)=-\frac{ \ell{1}(\frac{u-v}{2}-\eta) \ell{2}(\frac{u+v}{2}+\eta)}{\ell{1}(\eta ) \ell{2}(v+\eta)},\label{Beta}
\end{align}
one can verify that the set
$\left\{\ket{\d;n_1,n_2,\ldots,n_{M'}}_{I\!I}\right\}$ in (\ref{TwoSets}) is
exactly our chiral basis in (\ref{Basis;M}) when $M=0,1$ by letting
$r_1=u_0\pm 1$.

The difference between our set of generating states (\ref{Basis;M}) (see also Fig.~\ref{Fig1})
and the set $\left\{\ket{\d;n_1,n_2,\ldots,n_{M'}}_{I\!I}\right\}$ appears
when $M>1$.  For $M=2$, we have
\begin{align}
\ket{\d;n_1,n_2}_{I\!I}&=\bigotimes_{k_1=1}^{n_1-1}\psi'(r_{2d+k_1-1})\bigotimes \psi'(t_{2d+n_1-1})\bigotimes_{k_2=n_1+1}^{n_2-1}\psi'(r_{2\d+k_2-3})\bigotimes \psi'(t_{2\d+n_2-3}) \bigotimes_{k_3=n_2+1}^{N}\psi'(r_{2\d+k_3-5})\no\\
&=(-1)^N\beta_1(\rho_1,\varrho_1)\beta_2(\rho_2,\varrho_2)\underline{\bigotimes_{k_1=1}^{n_1}\psi'(r_{2d+k_1-1})\bigotimes_{k_2=n_1+1}^{n_2-1}\psi'(r_{2\d+k_2-3}) \bigotimes_{k_3=n_2}^{N}\psi'(r_{2\d+k_3-5})}\no\\[4pt]
&\quad+(-1)^N \beta_1(\rho_1,\varrho_1)\beta_1(\rho_2,\varrho_2)\ket{\d-1;n_1,n_2}+(-1)^N\beta_2(\rho_1,\varrho_1)\beta_1(\rho_2,\varrho_2)\ket{\d-1;n_1-1,n_2}\no\\[4pt]
&\quad+(-1)^N\beta_2(\rho_1,\varrho_1)\beta_2(\rho_2,\varrho_2)\ket{\d-1;n_1-1,n_2-1},\label{TwoBases}\\[2pt]
\rho_1&=t_{2d+n_1-1},\quad \varrho_1=r_{2d+n_1-3},\quad\rho_2=t_{2d+n_2-3},\quad \varrho_2=r_{2d+n_2-5},\no
\end{align}
where we let $r_1=u_0\pm1$. The underlined part in (\ref{TwoBases}) is beyond
our set of generating states in (\ref{Basis;M}) when $n_2=n_1+1$.  In this case, indeed, the
underlined part becomes
\begin{align}
&\qquad \underline{\bigotimes_{k_1=1}^{n_1}\psi'(r_{2d+k_1-1}) \bigotimes_{k_2=n_1+1}^{N}\psi'(r_{2\d+k_2-5})},\label{TwoBases1}
\end{align}
which corresponds to a \textit{double} drop of phase over a single link
$n_1,n_1+1$ (i.e.~two kinks on one link, in our notation) which is not
allowed in our basis, see Fig.~\ref{Fig1} and (\ref{Basis;M}).  For larger
$M>2$ the set $\left\{\ket{\d;n_1,n_2,\ldots,n_{M'}}_{I\!I}\right\}$, expanded
as in (\ref{TwoBases}), will produce states of type ``some double kinks plus
some ordinary kinks", while triple, quadruple etc.~kinks on one link will
never occur.  In addition, double kinks (double drop of phase) cannot occur
on consecutive links.

Consequently, the set $\left\{\ket{\d;n_1,n_2,\ldots,n_{M'}}_{I\!I}\right\}$
contains extra vectors when $M\geq 2$ so that Baxter's basis
(\ref{BaxterBasis2}) is a more general one.  The sets
$\{\ket{\d;n_1,n_2,\ldots,n_{M'}}_{I\!I}\}$ and
$\{\ket{\d;n_1,n_2,\ldots,n_{M'}}_{I\!I\!I}\}$ in basis (\ref{BaxterBasis2})
reduce respectively to our chiral basis in (\ref{Basis;M}) as follows
\begin{align}
&\{\ket{\d;n_1,\ldots,n_{M'}}\}_{I\!I}\to\{\ket{\d;n_1,n_2,\ldots,n_{M}}\}:\quad  v=\frac{\eta+r-t}{2}+1,\quad u_0=\frac{r+t}{2},\quad M=M',\no\\
&\{\ket{\d;n_1,\ldots,n_{M'}}\}_{I\!I\!I}\to\{\ket{\d;n_1,n_2,\ldots,n_{M}}\}:\quad  v=\frac{\eta+r-t}{2}+1,\quad u_0=\frac{r+t}{2}+\eta,\quad M=M'.\label{Reduce}
\end{align}

When $N=2M$, the basis (\ref{BaxterBasis2}) with generic $r$, $t$ and $v$ is
argued to be complete \cite{Baxter4}.  As a consequence, we can use it to
expand the remaining eigenstates beyond our chiral basis.  From our analysis in
Appendix \ref{AppendixE}, we conjecture that the remaining eigenstates ($M\geq
2$) correspond to the ``bound pair'' solutions of (\ref{Baxter;BAE})
\cite{Baxter4} with
\begin{align}
&\mu_1=\frac{\eta}{2},\quad\mu_2=-\frac{\eta}{2},\quad\frac{\ell{1}(\mu_1+\frac\eta 2)}{\ell{1}(\mu_1-\frac\eta 2)}\to \infty,\quad \frac{\ell{1}(\mu_2+\frac\eta 2)}{\ell{1}(\mu_2-\frac\eta 2)}\to 0,\quad \frac{\ell{1}(\mu_1+\frac\eta 2)}{\ell{1}(\mu_1-\frac\eta 2)}\frac{\ell{1}(\mu_2+\frac\eta 2)}{\ell{1}(\mu_2-\frac\eta 2)}=-1,\label{BoundPair1}\\
&\ell{1}(\mu_2-\mu_1+\eta)=-\ell{1}(2\eta)\varpi^{2}\,\left[\frac{\ell{1}(\mu_1+\frac{\eta}{2})}{\ell{1}(\mu_1-\frac{\eta}{2})}\right]^{-N}\prod_{k\neq 1,2}^{N/2}\frac{\ell{1}(\mu_k-\frac{3\eta}{2})}{\ell{1}(\mu_k+\frac{\eta}{2})}.\label{BoundPair2}
\end{align}
Now the expansion coefficients of the basis in (\ref{Baxter;BS1}) all vanish,
so that we need to extract the terms with slowest decay.  By substituting the
``bound pair'' solution (\ref{BoundPair1}), (\ref{BoundPair2}) into
Eqs.~(\ref{Baxter;BS1})-(\ref{Y;function}) and eliminating an overall factor,
the analytic expression of the remaining eigenstates can be derived.

For instance when $N=2M=4$, the remaining eigenstate in Tab.~\ref{N4;data} can
be written as
\begin{align}
\ket{\Psi}_{R}&=\sum_{\d=1}^{Q}\left[\frac{\ket{\d;1,2}_I }{\ell{2}(\omega_{\d})\ell{2}(\omega_{\d-2})}-\frac{\ket{\d;2,3}_I}{\ell{2}(\omega_{\d+1})\ell{2}(\omega_{\d-1})}+
\frac{\ket{\d;3,4}_I}{\ell{2}(\omega_{\d+2})\ell{2}(\omega_{\d})}-\frac{\ket{\d;1,4}_I}{\ell{2}(\omega_{\d-1})\ell{2}(\omega_{\d+1})}\right]\no\\
&\propto\ket{1,2}-\ket{2,3}+\ket{3,4}-\ket{1,4}.\label{MissingEigenstate}
\end{align}

\end{document}